\documentclass[useAMS,usenatbib,aastex]{mnras}

\usepackage{graphicx}
\usepackage{amssymb,amsmath}
\usepackage{natbib}

\newcommand{\elik}{{\mathbf K}}

\newcommand{\kv}{\kappa_{\rm Virial}}

\newcommand{\hu}{\hat{U}}
\newcommand{\hv}{\hat{V}}
\newcommand{\hw}{\hat{W}}
\newcommand{\hcin}{\hat{T}}
\newcommand{\hpi}{\hat{\Pi}}

\newcommand{\hjcin}{\hat{J}}
\newcommand{\hpsi}{\hat{\Psi}}
\newcommand{\hphi}{\hat{\Phi}}
\newcommand{\hper}{\hat{p}}

\newcommand{\hp}{\hat{P}}
\newcommand{\hpe}{\hat{p}_e}
\newcommand{\hrho}{\hat{\rho}}

\newcommand{\deh}{\delta \hat{H}}

\newcommand{\hm}{\hat{M}}

\newcommand{\dhfro}{\deh_F}
\newcommand{\hh}{\hat{H}}
\newcommand{\hr}{\hat{R}}
\newcommand{\hbz}{\hat{Z}}

\newcommand{\hrl}{{\hat{R}_l}}
\newcommand{\hrr}{{\hat{R}_r}}
\newcommand{\hrg}{{\hat{R}_\Gamma}}
\newcommand{\hzg}{{\hat{Z}_\Gamma}}
\newcommand{\qg}{{q_\Gamma}}
\newcommand{\hz}{\hat{z}}
\newcommand{\ha}{\hat{a}}
\newcommand{\hf}{\hat{f}}
\newcommand{\hrs}{\hat{r}}
\newcommand{\ho}{\hat{\Omega}}
\newcommand{\hzeta}{\hat{\zeta}}

\newcommand{\hbsur}{\hat{S}}

\newcommand{\nddr}{\delta_{\hr}}
\newcommand{\nddrr}{\delta^2_{\hr\hr}}
\newcommand{\nddzz}{\delta^2_{\hbz\hbz}}

\newcommand{\ddrrr}{\partial^3_{\hr\hr\hr}}

\newcommand{\ddrrrr}{\partial^4_{\hr\hr\hr\hr}}

\newcommand{\ddzzzz}{\partial^4_{\hbz\hbz\hbz\hbz}}

\begin{document}

\title[Toroidal figures of equilibrium]{Toroidal figures of equilibrium from a 2nd-order accurate, accelerated SCF-method with subgrid approach}
\author[J.-M. Hur\'e and F. Hersant]
{J.-M. Hur\'e$^{1,2}$\thanks{E-mail:jean-marc.hure@obs.u-bordeaux1.fr} and
F. Hersant$^{1,2}$\\
$^{1}$Univ. Bordeaux, LAB, UMR 5804, F-33615, Pessac, France\\
$^{2}$CNRS, LAB, UMR 5804, F-33615, Pessac, France}

\date{Received ??? / Accepted ???}
 
\pagerange{\pageref{firstpage}--\pageref{lastpage}} \pubyear{???}

\maketitle

\label{firstpage}

\begin{abstract}
We compute the structure of a self-gravitating torus with polytropic equation-of-state (EOS) rotating in an imposed centrifugal potential. The Poisson-solver is based on isotropic multigrid with optimal covering factor (fluid section-to-grid area ratio). We work at $2$nd-order in the grid resolution for both finite difference and quadrature schemes. For soft EOS (i.e. polytropic index $n \ge 1$), the underlying $2$nd-order is naturally recovered for Boundary Values (BVs) and any other integrated quantity sensitive to the mass density (mass, angular momentum, volume, Virial Parameter, etc.), i.e. errors vary with the number $N$ of nodes per direction as $\sim 1/N^2$. This is, however, not observed for purely geometrical quantities (surface area, meridional section area, volume), unless a subgrid approach is considered (i.e. boundary detection). Equilibrium sequences are also much better described, especially close to critical rotation. Yet another technical effort is required for hard EOS ($n < 1$), due to infinite mass density gradients at the fluid surface. We fix the problem by using kernel splitting. Finally, we propose an accelerated version of the SCF-algorithm based on a node-by-node pre-conditionning of the mass density at each step. The computing time is reduced by a factor $2$ typically, regardless of the polytropic index. There is a priori no obstacle to applying these results and techniques to ellipsoidal configurations and even to $3$D-configurations.

\end{abstract}


\section{Introduction}

The theory of figures of equilibrium is a broad and ancient  astrophysical subject, aiming at understanding the flatenning of the Earth and stars due to proper spin and tides \citep{ivory31,cl62,james63,clem67,chandra87}. While very complex structures and shapes are permitted, matter can basically attain two configurations, depending on angular momentum and rotation profile \citep{chandra73,hachisu86,horedttextbook2004}: an ellipsoidal shape or a toroidal shape at high momentum. This latter class is of great interest since disks and rings are seen in various astrophysical environments \citep{ba01}, around planets and satellites \citep{gi82}, around young and evolved stars \citep{wash96,du14}, around stellar and supermassive black holes \citep{rouan98,ki11,el12}. The subject is less documented \citep{kowal1885,dyson1893b,wong74,hachisu86,od03,petroff08,slany13}, probably for technical reasons (i.e. the strong deviation with respect to sphericity). For compact systems subject to rapid evolutions, the framework of general relativity is pertinent \citep{kuw88,toh90,nel92,nier94,akl98,cs03,stu09,font10,hk13}. Most investigations of self-gravitating tori are, however, performed in classical gravity. Configurations have been computed in details by \cite{hachisu86} and collaborators \citep[see also][]{eri78,hashi93,eri93,lss93,kley96,horedttextbook2004}. Despite restrictive assumptions (polytropic gas, symetries, etc.), there is a wide and rich collection of possible equilibria \citep{ansorg03}. This paper revisits the determination of single-body, toroidal configurations from the Self-Consistent-Field (SCF)-method. A special attention is paid to the convergence process leading to the physical solution and to the preservation of the accuracy order, which points are rarely discussed in the litterature \citep{od03}. Here, we work at $2$nd-order in the grid resolution. Following the pioneering work by \cite{lanza87}, the Poisson-solver is based on multigrid. Indirectly, the polytropic Equation-Of-State (EOS) plays a critical role. For instance, we show that the fluid volume | a key-quantity when building equilibrium sequences | appears quite uncertain due to curvature effects at the inner and outer edges. This is fixed by a subgrid approach. There is an additionnal diffculty for $n<1$ : the presence of infinite gradients of the mass density at the fluid surface makes standard quadrature schemes unprecise. This is solved by treating the underlying power-law behavior separately. Globally, the fluid boundary needs to be accounted for in order to fully recover the order of underlying schemes, i.e. $2$nd-order. The Virial Parameter is then a good indicator of the precision of the solution and integrated values. Finally, we propose a semi-empirical recipe to speed-up the SCF-loop by a factor up to $2$ typically. It is a mass density preconditionner.

The article is organized as follows. In Sect. \ref{sec:physmod}, we list the assumptions, review the equation set and define the enthalpy associated with the polytropic EOS. We briefly comment on the existence of solutions, and in particular how the index $n$ shapes the mass density profile at the fluid surface and below. We put the problem into dimensionless form and write down the Virial equation, associated kernels, and the Virial parameter. In Sect. \ref{sec:solving}, we express the three constants of the problem and recall the principle of the SCF-method, including self-normalisation (which permits to isolate a particular solution). We briefly outline the multigrid method used to solve the Poisson equation, and define the associated nested grids, from the coarsest level to the finest one. The formula for Boundary Values is given. Then comes the discretization of the Poisson equation and a short error analysis which gives the optimal multigrid level. We discuss how the polytropic hypothesis impacts on the accuracy of quadratures at $2$nd-order. Section \ref{sec:perf} is devoted to examples obtained under rigid rotation for both hard and soft EOS, with index $n \in [0.5,1,1.5]$, at a low resolution of $128 \times 128$. We show how the gravitational, centrifugal and internal energy densities are distributed within the fluid section. The effect of various parameters (number of nested grids, covering factor) on the convergence process is discussed. In Sect. \ref{sec:subgrid}, we describe the subgrid approach and kernel splitting which are necessary to go beyond the raw model. A few examples demonstrate its efficiency. In Sect. \ref{sec:convacc}, we show that the number of SCF-iterations can be reduced by an appropriate preconditionning of the mass density field inside the loop. This gain (a factor $2$ typically) directly concerns the computing time and is therefore important when performing high resolution models. The last section is devoted to concluding remarks.

\section{Physical model}
\label{sec:physmod}

\begin{figure}
\includegraphics[width=8.7cm,bb=0 0 478 337,clip=]{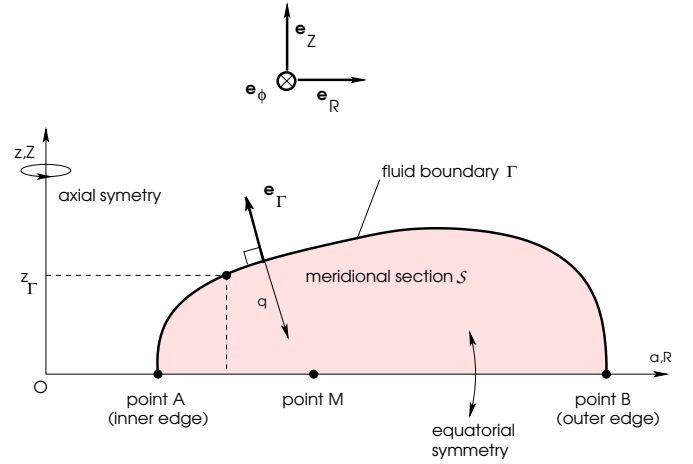}
\caption{Configuration for the self-gravitating torus (limited to the half-plane $Z \ge 0$). Points A and B are the radii at the inner and outer edges respectively, while point M is an interior point. The meridional section is denoted ${\cal S}$, $\vec{e}_\Gamma$ is the unit vector normal to $\Gamma$, the fluid surface.}
\label{fig:tore_et_ell}
\end{figure}

\subsection{Hypothesis}
\label{subsec:hypothesis}

We study the equilibrium of an isolated, compressible fluid with pressure $P$ and mass density $\rho$ rotating in its own gravitational potential $\Psi$ at rotation rate $\vec{\Omega}$. We assume no accretion, no internal circulation, and no viscosity. The Equation-Of-State (EOS) is of the form $f(P,\rho)=0$. Barotropes (and in particular polytropes) enable to cover various situations, e.g. perfect gas, gas with relativistic electrons, degenerate gas, convective media, etc. \citep{cox1968}. The analysis is performed in cylindrical coordinates $\vec{r}(R,\phi,Z)$ with vector basis $(\vec{e}_R,\vec{e}_\phi,\vec{e}_Z)$, and the rotation axis is along $\vec{e}_Z$, i.e. $\vec{\Omega}=\Omega \vec{e}_Z$.

The meridional section of the fluid is denoted ${\cal S}$, and it intercepts the fluid surface through a closed curve $\Gamma(\vec{r}_\Gamma)$. We define a unit vector $\vec{e}_\Gamma$ locally normal to $\Gamma$ and oriented outward and the associated coordinate is
\begin{equation}
q - q_\Gamma=(\vec{r}_\Gamma-\vec{r}) \cdot \vec{e}_\Gamma,
\end{equation}
with $q=q_\Gamma$ onto $\Gamma$ and $q>q_\Gamma$ inside the fluid (i.e. below the surface). Under axial and equatorial symmetries, we have
\begin{flalign}
\begin{cases}
\partial_\phi f = 0,\\
f(Z)-f(-Z)=0,
\end{cases}
\end{flalign}
for any scalar quantity $f$. The system and notations are summarized in Fig. \ref{fig:tore_et_ell}. We focus on toroidal configurations, where matter has deserted the rotation axis.

\subsection{Equation set}

According to the Schwarz condition, the angular momentum and subsequently the rotation rate are independant of the $Z$-coordinate \citep{amendt1989}. The circular velocity is of the form $\vec{v} = \Omega(R) R \vec{e}_\phi$ and the equilibrium of the self-gravitating fluid is described by the equation set
\begin{flalign}
\begin{cases}
 - \frac{\vec{v}^2}{R} \vec{e}_R=-\frac{1}{\rho} \vec{\nabla} P -\vec{\nabla} \Psi,\\
f(P,\rho)=0,\\
\Delta \Psi = 4 \pi G \rho.\\
\end{cases}
\label{eq:eset}
\end{flalign}
The Poisson equation, i.e. Eq.(\ref{eq:eset}c), is subject to boundary conditions. Boundary values (BVs) are accessible uniquely from the generalized Newton's law
\begin{equation}
\Psi=-G\iiint_\text{fluid}{\frac{\rho(\vec{r}') d^3r'}{|\vec{r}-\vec{r}'|}},
\label{eq:psinewton}
\end{equation}
where $\vec{r}$ and $\vec{r}'$ are spherical vectors. If we define the centrifugal potential $\Phi$ as
\begin{equation}
\Phi = -\int{\Omega^2(R)RdR} <0,
\label{eq:centpot}
\end{equation}
then Eq.(\ref{eq:eset}a) takes the form
\begin{flalign}
\frac{1}{\rho} \vec{\nabla} P + \vec{\nabla} (\Psi + \Phi)=0.
\label{eq:eset_bis}
\end{flalign}

The coupling between the rotation rate and the spatial distribution of matter is a tricky problem which exceeds the present context \citep[e.g.][]{binneytremaine87}. In \cite{balm92}, the link is made through conservation of potential vorticity. In general however, $\Omega$ remains an input profile. Various options are classically considered \citep{bo73,em85,hachisu86,eri93,reese06}, including power laws of the radius \citep[e.g.][]{slany13}. We will not favour a particular profile, except for illustrative purposes.

\subsection{Enthalpy field}

Equation (\ref{eq:eset_bis}) can be fully integrated provided the pressure term is the gradient of a scalar field $H$, the enthalpy, i.e.
\begin{equation}
 \frac{1}{\rho} \vec{\nabla} P \equiv \vec{\nabla} H,
\label{eq:hentgrad}
\end{equation}
which is possible for a polytropic EOS
\begin{equation}
P=K \rho^\gamma,
\end{equation}
where $K$ and $\gamma$ are positive constants. We thus have
\begin{equation}
H = K \frac{\gamma}{\gamma-1}\rho^{\gamma-1} + cst,
\label{eq:enthalpy}
\end{equation}
where the constant is generally forgotten. So, the equation set becomes
\begin{flalign}
\begin{cases}
\vec{\nabla} \left(\Phi+H+\Psi \right) = \vec{0},\\
H = K (n+1) \rho^{1/n},\\
\Delta \Psi = 4 \pi G \rho,\\
\end{cases}
\label{eq:eset2}
\end{flalign}
where the polytropic index $n \ge 0$ is defined by
\begin{equation}
\gamma=1+\frac{1}{n}.
\end{equation}

For isothermal structures where $\gamma=1$ and possibly for layered systems made of different polytropes where continuity is required \citep{singh70,mf85,ca86,bee88,ru88,cmc00,horedttextbook2004}, Eq.(\ref{eq:enthalpy}) still holds but the integration constant must be accounted for. Actually, with the definition \citep{kimura81}
\begin{flalign}
H = K\frac{\gamma}{\gamma-1} \left(\rho^{\gamma-1} - \rho_0^{\gamma-1} \right),
\label{eq:enthalpy_iso}
\end{flalign}
where $\rho_0$ is a constant, we have
\begin{equation}
\lim_{\gamma  \rightarrow 1} H = K \ln \frac{\rho}{\rho_0}.
\end{equation}

The incompressible approximation $n =0$ is also of great interest \citep{chandra73}. In this case, Eq. (\ref{eq:eset2}b) is obsolete because $\rho$ is uniform in the system throughout and disconnected from the enthalpy field.

\begin{figure}
\includegraphics[width=8.3cm,bb=0 0 426 363,clip==]{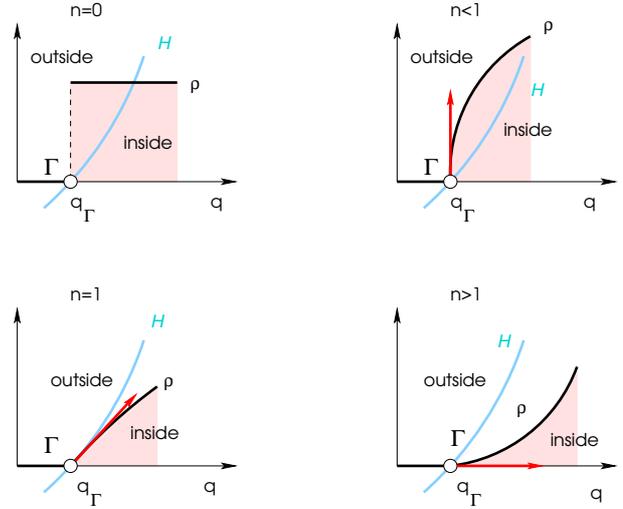}
\caption{Expected density profile ({\it bold}) and derivative ({\it red arrow}) when crossing the fluid boundary $\Gamma$ normally (coordinate $q$), depending on the polytropic index $n$ (see also Fig. \ref{fig:tore_et_ell}). The enthalpy is shown ({\it blue}).}
\label{fig:gradrho}
\end{figure}

\subsection{Comment on the existence of solutions}
\label{subsec:fb}

We see from Eq.(\ref{eq:eset2}a) that $H + \Psi + \Phi$ is a space invariant, namely
\begin{equation}
H = C - \Psi - \Phi,
\label{eq:hcontinuous}
\end{equation}
where $C$ is a constant. Because $\Psi$ and $\Phi$ are both negative, and due to the correlation between $\psi$ and $\rho$ (the larger the mass density, the deeper the potential well), $H$ is expected to rise inside matter, with one maximum (or more). The regions occupied by the fluid (where $\rho>0$) are didacted by the value of constant $C$ (to be compared with Jacobi's constant in the restricted three body problem). When $C>0$, matter fills the whole physical space. So, finite size systems require $C<0$.

Clearly, Eq. (\ref{eq:eset2}b) allows for a large variety of mathematical solutions, depending on the polytropic index. For some indices, $\rho$ can be negative or complex. We retain the physical solutions associated with $H>0$ for any $n$. This condition is explicitly stated by replacing Eq.(\ref{eq:eset2}b) by
\begin{equation}
K(n+1)\rho^{1/n} = \sup \left(H,0\right).
\label{eq:rhoh_sup}
\end{equation}
This cut-off is a major difference with the approach based on the Lane-Emden equation, apart from the rotation term. Besides, we focus on single bodies while $H>0$ can occur in two disconnected regions (or more), leading to multibody configurations \citep[see e.g.][]{hen86}.

\subsection{Fluid surface and interior}
\label{subsec:fsi}

If the fluid is isolated, the surface is defined by $\rho=0$. This corresponds to the zero-enthalpy level. In the isothermal limit (and possibly for other particular indices), this surface may not exist at all, or be rejected at infinity, which case is well known for spherical configurations \citep{sch90,horedttextbook2004}. Just below the boundary, $\rho$ rises according to Eq.(\ref{eq:rhoh_sup}). Perpendicularly to $\Gamma$ (see Sec. \ref{subsec:hypothesis} and Fig. \ref{fig:tore_et_ell}), the mass density gradient is 
\begin{equation}
\nabla_q \rho \sim n \nabla_q H  \times \rho^\frac{n-1}{n},
\end{equation}
which, by linearly expanding $H$ around $q=q_\Gamma$, is given by
\begin{equation}
\nabla_q \rho \sim n(q-q_\Gamma)^{n-1}.
\end{equation}
As a result, the mass density gradient is finite at the fluid boundary for $n \ge 1$ (soft EOS), but infinite for $n < 1$ (hard EOS). This latter case introduces a difficulty from a numerical point of view (see below). Since the gravitational potential is a continuous function of space \citep{kellogg29,durand64} as well as the rotation law, $H$ is also continuous, which property is transmitted to $\rho$ inside the fluid, from Eq.(\ref{eq:hcontinuous}). Figure \ref{fig:gradrho} summarizes four different situations : the larger the index, the more peaked the mass density, and the wider the wings. Conversely, the closer to zero the index, the sharper the transition from the fluid to the external medium, with a bare jump for $n=0$.

\subsection{Dimensionless problem}

It is convenient to work with adimensionned quantities, especially for a numerical approach. We use the generic notation $f_0 \hf \equiv f$ where $f_0$ is the magnitude of $f$, which leaves $\hf$ of the order of unity. Space coordinates are normalized by the physical length scale $L$, i.e. 
\begin{flalign}
\begin{cases}
\hr\equiv \frac{R}{L},\\
\hbz\equiv \frac{Z}{L},
\end{cases}
\end{flalign}
and we set
\begin{flalign}
\begin{cases}
\hrho \equiv \frac{\rho}{\rho_0},\\
\hp \equiv \frac{P}{P_0},\\
\ho  = \frac{\Omega}{\Omega_0}\\
\hpsi \equiv \frac{\Psi}{\Psi_0},\\
\hh \equiv \frac{H}{H_0},\\
\hphi \equiv \frac{\Phi}{\Phi_0},
\end{cases}
\end{flalign}
for the mass density, pressure, rotation rate, gravitational potential, enthalpy and centrifugal potential respectively, where
\begin{flalign}
\begin{cases}
P_0=K \rho_0^\gamma,\\
\Psi_0 = G\rho_0L^2,\\
H_0=K(n+1)\rho_0^{1/n},\\
\Phi_0=\Omega_0 L^2.
\end{cases}
\end{flalign}
The dimensionless equation set then writes
\begin{flalign}
\begin{cases}
\hpsi + C_1\hh + C_2 \hphi = C_3,\\
\hrho^{1/n}= \sup(\hh,0),\\
\Delta \hpsi = 4 \pi \hrho,\\
\end{cases}
\label{eq:eset3}
\end{flalign}
where
\begin{flalign}
\hphi = -\int{\ho^2(\hr)\hr d\hr},
\label{eq:phidedims}
\end{flalign}
\begin{equation}
\hpsi=-\iiint_\text{fluid}{\frac{\hrho(\vec{\hrs}')}{|\vec{\hrs}-\vec{\hrs}'|} d^3\hrs'},
\label{eq:psinewton_adim}
\end{equation}
for BVs, and the constants are
\begin{flalign}
\begin{cases}
C_1\equiv \frac{ K (n+1)\rho_0^{\gamma-2}}{GL^2},\\
C_2\equiv \frac{\Omega^2_0}{G\rho_0},
\end{cases}
\label{eq:c1c2c3}
\end{flalign}
for finite indices. For $\gamma=1$, we have
\begin{flalign}
\begin{cases}
H_0=K,\\
\hrho=e^{\hh},\\
C_1=\frac{K}{G\rho_0 L^2}.
\end{cases}
\end{flalign}
While notations and definitions can vary from one author to another, these formulae are standard. Note that constant $C_1$ does not exist in \cite{hachisu86}, since it is included into the definition of $\hh$.

\subsection{Virial theorem and associated test}
\label{subsec:virial}

The Virial theorem is established by projecting the vectorial Euler equation, i.e. Eq.(\ref{eq:eset}a), on the position-vector $ \vec{r}'$ and integrating over the mass distribution. It is an energy equation which shows how the gravitational $W$, internal $U$ and kinetic $T$ energies are spatially distributed \citep[e.g.][]{cox1968}. For the actual problem where only the gravitational and pressure forces are present, the theorem reads
\begin{equation}
W+2T+U=0,
\label{eq:viriel}
\end{equation}
where
\begin{flalign}
\begin{cases}
W=\frac{1}{2} \iiint{\Psi dm},\\
T=\frac{1}{2}\iiint{ v^2 dm},\\
U=3 \Pi,\\
\Pi= \iiint{\frac{P}{\rho}dm}.\\
\end{cases}
\end{flalign}
When other forces are acting in volume or at the surface, the Virial equation must be modified. This is for instance the case for loaded polytropes \citep{fxw83}, with magnetic fields \citep{ye06}, when the fluid is overpressurized \citep{ui86}, for multibody configurations \citep{tn00}, etc. Given axial symmetry and Eq.(\ref{eq:c1c2c3}), we easily show that Eq.(\ref{eq:viriel}) becomes
\begin{equation}
\hw+\frac{C_1}{n+1}\hu+2 C_2 \hcin=0,
\label{eq:viriel_dedim}
\end{equation}
for finite indices $n$ (the second term is just $C_1\hu$ for $\gamma=1$), where
\begin{flalign}
\begin{cases}
W = G L^5 \rho_0^2 \hw,\\
\Pi= P_0 L^3 \hpi,\\
T= L^5 \rho_0 \ho_0^2  \hcin,
\end{cases}
\label{eq:wth}
\end{flalign}
and
\begin{equation}
\begin{cases}
\hw=\pi\iint_{{\cal S}}{\hpsi \hrho \ha d\ha d\hz},\\
\hu=3 \hpi,\\
\hpi=2\pi \iint_{{\cal S}}{\hp \ha d\ha d\hz},\\
\hcin=\pi\iint_{{\cal S}}{\ho^2  \hrho \ha^3 d\ha d\hz}.
\end{cases}
\label{eq:wth_dedim}
\end{equation}

Interestingly enough, Eq.(\ref{eq:viriel_dedim}) can be written in the form
\begin{equation}
2 \pi \iint_{{\cal S}}{\kv d\ha d\hz}=0,
\label{eq:intkappavir}
\end{equation}
where
\begin{equation}
\kv = \kappa_W + 2\kappa_T+\kappa_U,
\end{equation}
is the energy per unit meridional section area and
\begin{equation}
\begin{cases}
\kappa_W \equiv  \frac{1}{2} \hrho \hpsi \ha,\\
\kappa_U \equiv \frac{3C_1}{n+1} \hp \ha,\\
\kappa_T \equiv  \frac{1}{2} C_2 \ho^2  \hrho \ha^3,
\end{cases}
\end{equation}
with $\kappa_U = 3C_1 \hp \ha$ in the isothermal case. This kernel varies in space: it is a positive where dispersive forces dominate, and negative when cohesion (i.e. gravitational) forces dominate.

When Eq.(\ref{eq:eset3}) is solved numerically, Eq.(\ref{eq:viriel_dedim}) is expected to be satisfied too. Since numerical methods have not infinite precision but produce small errors, the Virial equation is generally not stricly zero. The self-consistency of a solution can then be checked a posteriori by comparing the left-hand-side of Eq.(\ref{eq:viriel_dedim}) to its largest term. Relative to the gravitational term, the Virial test writes
\begin{equation}
VP\overset{?}{=}0,
\end{equation}
where the Virial parameter is
\begin{equation}
VP = \frac{1}{|\hw|} \left(\frac{C_1}{n+1}\hu+2 C_2 \hcin \right)-1.
\end{equation}

This test does not prove the exactness of the computed solution, but only its self-consistency. It must also be interpreted with caution since, in numerical calculus, tiny quantities are difficult to determine with precision.

\section{Solving the non-linear problem at $2$nd-order}
\label{sec:solving}
 
We see that $\hrho$ (or $\hh$) is the solution of a non-linear, integro-algebraic equation. Given $L$, $\rho_0$, $\Omega_0$, $K$ and $n$, one can compute $C_1$ and $C_2$ from Eq.(\ref{eq:c1c2c3}), but constant $C_3$ remains undertermined unless $\hpsi$, $\hh$ and $\hrho$ are known at some point in space. On the contrary, if the three constants are set arbitrarily, the solution in the form of $\hpsi$, $\hh$ and $\hrho$ may not exist. Actually, for three reference points A, B, and M of space, these constants must satisfy
\begin{flalign}
\begin{cases}
\hpsi_A + C_1\hh_A + C_2 \hphi_A = C_3,\\
\hpsi_B + C_1\hh_B + C_2 \hphi_B = C_3,\\
\hpsi_M + C_1\hh_M + C_2 \hphi_M = C_3,
\label{eq:bernouilliabc}
\end{cases}
\end{flalign}
which is not automatic. By selecting points A and B onto $\Gamma$ where the enthalpy is zero, we get
\begin{flalign}
\begin{cases}
C_1=-\frac{\hpsi_M\Delta \hphi_{AB}+\hpsi_A\Delta \hphi_{BM}+\hpsi_B\Delta \hphi_{MA}}{\hh_M\Delta \hphi_{AB}},\\
C_2=-\frac{\Delta \hpsi_{AB}}{\Delta \hphi_{AB}},\\
C_3= \frac{\hpsi_B  \hphi_A - \hphi_B \hpsi_A}{\Delta \hphi_{AB}},
\end{cases}
\label{eq:ctebc_hahbnull}
\end{flalign}
with $\Delta \hphi_{AB}=\hphi_A-\hphi_B$ and so on for $\Delta \hphi_{BM}$ and $\Delta \hphi_{MA}$.

\subsection{Principle of the SCF-method}
\label{subsec:scfloop}

 To get the equilibrium, the three constants are often regarded as unknowns and solved together with the enthalpy, mass density and gravitational potential through an iterative procedure. A common method consists in computing successively : i) the mass density from the enthalpy through Eq.(\ref{eq:eset3}b), ii) the gravitation potential from Eq.(\ref{eq:eset3}c) and the centrifugal potential from Eq.(\ref{eq:phidedims}), iii) constants $C_1$, $C_2$ and $C_3$ from Eqs.(\ref{eq:ctebc_hahbnull}), and iv) the enthalpy from Eq.(\ref{eq:eset3}a), and so on until input and output match. This is the so-called ``Self-Consistent Field''-(SCF) method \citep{om68,hachisu86}. It requires a starting guess $\hh^{(0)}$ and a convergence criterion to detect, at some step $t$ within the cycle, if the enthalpy $\hh^{(t)}$ has converged in the numerical sense. The capacities of convergence of the SCF-method are seemingly misunderstood \citep{od03}, but it happens to work in most cases if : i) the $3$ reference points are conveniently chosen, and ii) the enthalpy is appropriately rescaled within the cycle. Traditionnally, points A and B are radii at the inner and outer edge of the torus respectively. Since $\ho$ is generally monotonic with the radius, this ensures that $\Delta \hphi_{AB}$ is maximum.

\subsection{Enthalpy normalisation}
\label{subsec:Mrecsaling}

As outlined in \cite{hachisu86}, $\hh$ does not necessarily remain under control. From Eqs.(\ref{eq:eset3}a) and (\ref{eq:eset3}b), we have
\begin{equation}
\hrho^n = \frac{C_3-C_2 \hphi-\hpsi(\hrho)}{C_1},
\end{equation}
and we see that both sides respond in the same way when $\hrho$ varies. The SCF-cycle is therefore prone to runaway. However, since neither $C_2$ nor $C_3$ depend on point M, we have an extra degree of freedom to avoid any drift. We can impose that the enthalpy at some point in the fluid takes a particular value (see Sect. \ref{subsec:Mrecsaling}). By inserting $C_1^{(t)}$ into Eq.(\ref{eq:scfnewh}), we get
\begin{equation}
\hh^{(t+1)} = \hh_M^{(t)} \times \frac{C_3^{(t)}-C_2^{(t)} \hphi^{(t)}-\hpsi^{(t)}}{C_3^{(t)}-C_2^{(t)} \hphi_M^{(t)}-\hpsi_M^{(t)}},
\label{eq:scfnewh_v2}
\end{equation}
meaning that if M is held fixed, its enthalpy is constant during the SCF-iterations. Then, at convergence : i) $\hh_M$ has conserved its initial value, and ii) the maximum enthalpy is not under control, and it does not necessarily occur at point M. This is not an obstacle to convergence, but there are clearly an infinity of solutions, each dictated by $\hh_M^{(0)}$. We follow \cite{hachisu86} by requiring
\begin{flalign}
\begin{cases}
\hh_M=\max(\hh),\\
\hh_M=1,
\end{cases}
\label{eq:hselfscaled}
\end{flalign}
at convergence. Now, point M is not known in advance but changes during the SCF-iterations. In fact, this ``self-normalized'' version is obtained by computing constants $C_2$ and $C_3$ at time $t$ and constant $C_1$ at time $t+1$ from the new estimate $\hh^{(t+1)}$. This is feasible because only $C_1$ depends upon $\hh_M$. At step $t$, we successively determine
\begin{enumerate}
\item $\hrho^{(t)}$ from Eq.(\ref{eq:eset3}b),
\item $\hpsi^{(t)}$ from  Eq.(\ref{eq:eset3}c),
\item $\hphi^{(t)}$ from Eq.(\ref{eq:phidedims}),
\item constants $C_2^{(t)}$ and $C_3^{(t)}$ from Eqs.(\ref{eq:ctebc_hahbnull}b) and (\ref{eq:ctebc_hahbnull}c),
\item $(C_1\hh)^{(t+1)}$ from Eq.(\ref{eq:eset3}a), namely
\begin{equation}
(C_1\hh)^{(t+1)} = C_3^{(t)}-C_2^{(t)} \hphi^{(t)}-\hpsi^{(t)},
\label{eq:scfnewh_hachisu}
\end{equation}
\item its maximum value, happening at some point M
\begin{equation}
\max\left[C_3^{(t)}-C_2^{(t)} \hphi^{(t)}-\hpsi^{(t)}\right] \equiv C_1^{(t+1)},
\end{equation}
\item and the new enthalpy field
\begin{equation}
\hh^{(t+1)} = \frac{C_3^{(t)}-C_2^{(t)} \hphi^{(t)}-\hpsi^{(t)}}{C_1^{(t+1)}}.
\label{eq:scfnewh}
\end{equation}
\end{enumerate}

Since the computational grid is fixed, $\hphi$ does not vary during the cycle and so step (iii) can be removed from the list and executed once for all. Convergence is checked after step (vii). A natural indicator of convergence is (see below):
\begin{equation}
\delta \hh^{(t+1)} = \hh^{(t+1)}-\hh^{(t)}.
\end{equation}

\begin{figure}
\includegraphics[width=8.7cm,bb=0 0 567 450,clip=]{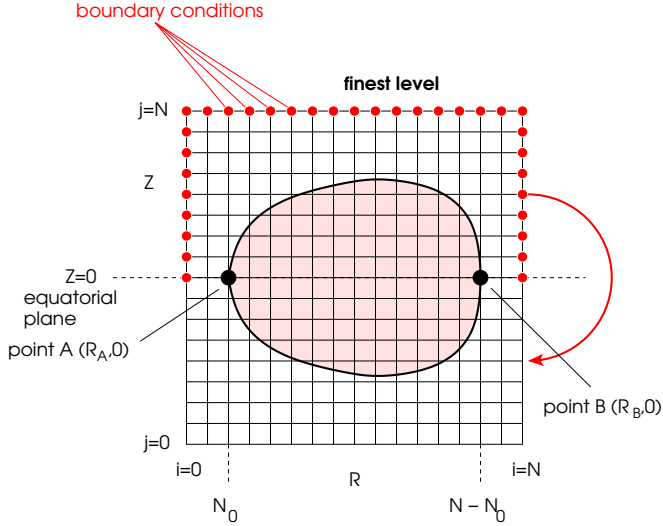}
\caption{Uniform grid used to solved the Poisson equation from multigrid. At the finest level, it contains $2^\ell+1$ nodes per direction. BVs are computed for part of the grid where $Z \ge 0$ and then copied below the equatorial plane. There are $N_0 \ge 0$ nodes leftward/rightward to the inner/outer edge free of matter.}
\label{fig:grid.eps}
\end{figure}

\subsection{Multigrid}
\label{subsec:grid}

The main difficulty arising in the actual problem is the determination of the potential in Eq.(\ref{eq:eset3}a) which varies during the SCF-iterations due to the evolution of the enthalpy. We solve Eq.(\ref{eq:eset3}c) for $\hpsi$ on an uniform grid $\{\hr_i,\hbz_j\}$ from multigrid. This technique is very efficient for solving partial differential equations (PDEs) since it combines nested grids and relaxation \citep[e.g.][]{Briggs00,gsg04,gt11}. It has already been employed in the context of figures of equilibrium \citep{lanza87,bl88,lanza92}. Short wavelengths are fastly smoothed by relaxation on a given grid, in contrast to long wavelengths. This is circumvented by using low resolution versions of the finest grid, thereby converting long wavelengths into small ones. The communication between all levels, from the coarsest to the finest grid and vice-versa, is achieved from restriction and prolongation operators, enabling to write fields and solve the discretized PDE at all levels. The restriction operator is based on full-weighting, while the prolongation operator employs bilinear interpolation.  All the grids are visited, from finest grid to coarsest and back. This is called a ``V-cycle''. At each level, one red-black Gauss-Seidel algorithm (or more) is performed. Despite a good convergence rate (about $1$ extra digit fixed by V-cycle), a unique V-cycle does not deliver the exact solution of the discretized problem, only a better approximation. However, by applying a V-cycle several times, multigrid becomes almost equivalent to matrix diagonalisation methods. In single precision, about $7$ successive V-cycles are necessary typically, and this is $15$ in double precision.

 Because multigrid has nominal performance when the spacing is the same in all directions, we use a square grid $[\hr_0,\hr_N] \times [\hbz_0,\hbz_N]$. In order to vary the fluid section-to-grid area ratio (i.e. the covering factor), we put $N_0 \ge 0$ nodes in between the inner and outer edge of the fluid and the grid boundary, as indicated in Fig. \ref{fig:grid.eps}. Since there are no forces susceptible to thicken the torus in the $Z$-direction (i.e. the section ${\cal S}$ is expected to be oblate), the grid size is imposed by the the equatorial diameter of the torus $2\hr_e=\hr_B-\hr_A$ (expected to be larger than the polar one $2\hr_p$). In spite of the equatorial symmetry, we found no real advantage to work in the half-plane $Z \ge 0$ only. This would mean i) Neumann conditions at the grid bottom $Z = 0$, ii) much more Dirichlet boundary values to compute, and iii) a smaller covering factor. The grid is therefore centered in the equatorial plane. Under these conditions, the number of subintervals between points A and B is
\begin{equation}
N' = N-2N_0,
\label{eq:nprim}
\end{equation}
and the numerical resolution of the finest grid is
\begin{equation}
\hat{h} = \frac{2\hr_e}{N'},
\label{eq:gres}
\end{equation}
both in the $\hr$- and $\hbz$-directions. The grid extends radially from $R_0$ to $R_N$ with
\begin{flalign}
\begin{cases}
\hr_0= \hr_A - N_0 \hat{h} \ge 0,\\
\hr_N= \hr_B + N_0 \hat{h},
\end{cases}
\label{eq:rorn}
\end{flalign}
and we have
\begin{flalign}
\begin{cases}
\hbz_N= \frac{1}{2}(\hr_N-\hr_0),\\
\hbz_0= -\hbz_N,
\end{cases}
\end{flalign}
in the vertical direction. The sampling is
\begin{flalign}
\begin{cases}
\hr_i=\hr_0+i\hat{h}, \quad i \in [0,N],\\
\hbz_i=\hbz_0+j\hat{h}, \quad j \in [0,N].
\end{cases}
\end{flalign}
This is for the finest level $\ell$ (i.e. highest resolution) and we have
\begin{equation}
N+1=2^\ell+1,
\label{eq:nl}
\end{equation}
per direction. The nested grids are easily generated. With mesh size doubling, the coarser grid at level $\ell-1$ is obtained by keeping nodes $2i$ and $2j$. At level $\ell-2$, we keep nodes $4i$ and $4j$. At level $1<k<\ell$, we keep nodes $2^{(\ell-k)}i$ and $2^{(\ell-k)}j$. For the coarsest level, $k=1$, there are just $3$ nodes per direction, $9$ in total, where $8$ belong to the grid boundary.

\subsection{Discretization of the Poisson equation. Error analysis}
\label{subsec:secondorderpoisson}

The discretization of Eq.(\ref{eq:eset3}c) writes
\begin{equation}
\left(\nddrr + \frac{1}{\hr_i} \nddr +  \nddzz \right) \hpsi_{i,j} + \tau_{i,j} = 4\pi \hrho_{i,j},
\label{eq:peqincyl_uvwithtau}
\end{equation}
where $\delta$ and $\delta^2$ denotes the approximations for the first and second derivatives and $\tau= {\cal O}(\hat{h}^2)$ is the truncation error. By using second-order centered schemes, the dimensionless potential at node $i,j$ is given by
\begin{flalign}
\nonumber
\hpsi_{i,j}=&\frac{1}{4}\left(a_+ \hpsi_{i+1,j} + a_- \hpsi_{i-1,j}\right.\\
&\qquad\left. + \hpsi_{i,j+1} + \hpsi_{i,j-1} - 4\pi \hrho_{i,j}\hat{h}^2\right),
\label{eq:poissonorder2}
\end{flalign}
where
\begin{equation}
\begin{cases}
a_+ = 1 + \frac{\hat{h}}{2\hr_i},\\
a_- = 1 - \frac{\hat{h}}{2\hr_i}.\\
\end{cases}
\end{equation}

The magnitude of the scheme error is
\begin{equation}
\tau \approx -\frac{\hat{h}^2}{12} \left( \ddrrrr + 2\frac{1}{R} \ddrrr +\ddzzzz \right) \hpsi,
\label{eq:taucorrection}
\end{equation}
and explicitly decreases with $\hat{h}$. As usual in error analysis \citep{opac-b1132370}, we assume that the third and fourth derivatives are bounded, i.e.
\begin{equation}
\left| \ddrrrr  \hpsi \right| + 2\frac{1}{R} \left|\ddrrr  \hpsi \right| + \left|\ddzzzz  \hpsi \right| \le D.
\label{eq:taucorrection}
\end{equation}
At a node $(i,j)$, the total error is then given by
\begin{equation}
E_{i,j}(\hat{h}) = \hpsi_{i,j}\left(\frac{4e}{\hat{h}^2} + \frac{e}{\hat{h} \hr} \right) +  \frac{\hat{h}^2}{12} D,
\label{eq:eij}
\end{equation}
where $e$ is the computer precision. As it is well known, finite differences become inaccurate if the spacing is too small, leaving round-off errors and loss of significance. The error is minimum for
\begin{equation}
\hat{h}^4-\frac{6e\hpsi_{i,j}}{D \hr}\hat{h} - \frac{48e\hpsi_{i,j}}{D} = 0.
\end{equation}
When the second derivative is the limiting quantity, the optimal spacing is $\hat{h}_{\rm opt} \sim e^{1/4}$. The first derivative can bring a similar contribution if $\hat{h}\sim \hr$ typically (i.e. the computational grid is close to the $Z$-axis). It all depends on $\hpsi/D$ which is difficult to estimate. The optimal spacing is in the range
\begin{equation}
 \left(\frac{6e\hpsi_{i,j}}{D \hr}\right)^{1/3} \lesssim \hat{h}_{\rm opt}  \lesssim  \left(\frac{48e\hpsi_{i,j}}{D}\right)^{1/4},
\end{equation}
which can be converted into an optimal level $\ell_{\rm opt}$ of the multigrid from Eq.(\ref{eq:nl}). With standard numbers, we get values listed in Table \ref{tab:ehl}. 

\begin{table}
\begin{tabular}{lcc}
          & single precision            & double precision \\ \hline
$e$                & $6 \times 10 ^{-8}$ & $1 \times 10 ^{-16}$ \\
 $\hat{h}_{\rm opt}$ &$0.006-0.04$        & $3 \times 10^{-4} - 9 \times 10^{-6}$ \\
 $\ell_{\rm opt}$   & $5-7$              & $12-17$\\
 $\frac{E}{\hpsi}$  & $0.007- 2 \times 10^{-4}$     & $5 \times 10^{-9} -  5\times10^{-6}$ \\\\               
 $\ell$            &                               & $7$ \\
absolute error  &      & $\sim \hat{h}^2 \sim 2 \times 10^{-5}$ \\ \hline
\end{tabular}
\caption{Optimal spacing and associated multigrid level for $2$nd-order accuracy, depending on the computer precision $e$. The last rows give the absolute error expected in potential values for $\ell=7$.}
\label{tab:ehl}
\end{table}

\subsection{Boundary values (BVs)}
\label{subsec:bcs}

Under axial symmetry and equatorial symmetries, Eq.(\ref{eq:psinewton}) simplifies somehow. The adimensionned potential takes the form \citep{durand64,hure05}
\begin{equation}
\hpsi =\iint_{\cal S}{\hrho \kappa d\hz d\ha},
\label{eq:psibcs}
\end{equation}
where
\begin{flalign}
\begin{cases}
\ha\equiv \frac{a}{L},\\
\hz\equiv \frac{z}{L},
\end{cases}
\end{flalign}
\begin{equation}
\kappa=  -2 \sqrt{\frac{\ha}{\hr}} \left[k_+\elik(k_+)+k_-\elik(k_-)\right],
\label{eq:kappa}
\end{equation}
\begin{equation}
\elik(k) = \int_0^{\pi/2}{\frac{d\phi}{\sqrt{1-k^2 \sin^2 \phi}}}
\end{equation}
is the complete elliptic integral of the first kind,
\begin{equation}
k_\pm=\frac{2\sqrt{\ha\hr}}{\sqrt{(\ha+\hr)^2+\hzeta_\pm^2}}
\label{eq:kmod}
\end{equation}
is the modulus, $\hzeta_+=\hbz-\hz$ and $\hzeta_-=\hbz+\hz$. When $k \rightarrow 1$ (which occurs when $\hr \rightarrow \ha$ and $\hzeta \rightarrow 0$), $\kappa$ diverges logarithmically, and the determination of $\Psi$ can become inaccurate. This never happens here since the fluid and the grid boundary are never in contact as soon as $N_0 > 0$ (see Sect. \ref{subsec:grid}).

\subsection{Note on quadratures}
\label{subsec:quad}

We need a quadrature scheme that preserves the $2$nd-order of the finite different scheme. Not only BVs are concerned, but also most output quantities (mass, volume, angular momentum, Virial parameter, etc.). The trapezoidal rule is the natural choice. For any function $\hf$ that vanishes outside the fluid (if it depends on $\rho$), we can consider the full grid, i.e.
\begin{flalign}
\label{eq:trap}
\frac{1}{\hat{h}^2} \iint_{\rm fluid}{\hf d \ha d \hz} &\approx \sum_{i \in [0,N],j \in [0,N]}{\hf_{i,j}}\\
& \nonumber  \qquad -\frac{1}{2}\sum_{i \in [0,N]}{(\hf_{i,0}+\hf_{i,N})}\\
&\nonumber  \qquad -\frac{1}{2}\sum_{j \in [0,N]}{(\hf_{0,j}+\hf_{N,j})}\\
&\nonumber  \qquad + \frac{1}{4}\left(\hf_{0,0} +\hf_{N,0} + \hf_{0,N} + \hf_{N,N}\right).
\end{flalign}
For geometrical quantities like the fluid volume, one must multiply $\hf$ by a weighting function $w$, namely
\begin{equation}
w=
\begin{cases}
1 & \text{if} \quad \hh\ge0 \quad \text{(inside the fluid)},\\
0 & \text{if} \quad \hh<0 \quad \text{(outside)}.
\end{cases}
\end{equation}

There are two pending problems that can damage the $2$nd-order. First, $\Gamma$ is rounded and does not follow the grid lines (see Fig. \ref{fig:tore_et_ell}). Standard quadrature schemes are generally not well suited to irregular or curved integration domains. This is a natural motivation for space mapping \citep{gc01,2016arXiv160502359R}. Second, for hard EOS ($n < 1$), $\hrho$ does not reach $\Gamma$ through a finite gradient. Again, this is a situation where many quadrature schemes fail to be accurate. The question of determining the fluid boundary $\Gamma$ is then naturally addressed.

It is instructive to estimate the relative contribution of the numerical cells located just below the fluid surface. Let $\hf_\Gamma$ denote the typical value of $\hf$ onto or close to $\Gamma$, $\langle \hf \rangle$ be the mean value of $\hf$ inside the fluid, $\hper$ the perimeter and $\hbsur$ the fluid section. This contribution is given by
\begin{equation}
\frac{\hf_\Gamma  \hper \hat{h}}{\iint_{\rm fluid}{\hf d\ha d\hz}} \sim \frac{\hf_\Gamma  \hper}{\langle \hf \rangle \hbsur} \hat{h}.
\label{eq:fgamma}
\end{equation}
It is first-order in $\hat{h}$ if $\hf_\Gamma/\langle \hf \rangle \sim 1$, but second-order if $\hf_\Gamma/\langle \hf \rangle \sim \hat{h}$. Further, if $\hf$ depends linearly on $\hrho$ (see Sect. \ref{subsec:fsi}), then this contribution is $\sim \hat{h}^{1+n}$. It immediately follows that all geometrical quantities like the fluid volume are only $1$st-order accurate, and all quantities which are senstive to the mass density (BVs, mass, angular momentum, Virial kernels, etc.) are $2$nd-order accurate as soon as $n \ge 1$ (i.e. for soft EOS). For $n<1$, the situation is worse (see Sec. \ref{subsec:fb}): not only the fluid boundary is first-order, but the gradient of mass density is infinite at $\Gamma$. Increasing the order of the quadrature scheme is not the right way to fix the problem. Instead, a specific treatment must be considered. This is discussed in Sect. \ref{sec:subgrid}.

\section{Performances at $2$nd-order}
\label{sec:perf}

We have built a Fortran 90 code nammed {\tt DROP} for ``Differentially ROtating Polytropes'' for solving toroidal figures of equilibrium by using methods described above. The main input parameters are (see Sec. \ref{subsec:grid}) 
\begin{itemize}
\item $\ell$ (or $N$) for the multigrid method and grid,
\item $n$, $\hr_A$, $\hr_B$ for the polytropic torus,
\item $N_0$ for the covering factor.
\end{itemize}

The starting guess $\hh^{(0)}$ is obviously a critical point. We use a paraboloid with circular section (radius $\hr_e$) which is, in general, a good choice. Regarding the convergence criterion, we use the Frobenius norm $\deh_F$, namely
\begin{flalign}
\deh_F^2 \equiv \frac{1}{(N-1)^2}\sum_{i,j=1,N-1}{\deh_{i,j}^2},
\end{flalign}
which is a good compromise, and so the solution is found when
\begin{equation}
\deh_F \le \epsilon,
\label{eq:condh}
\end{equation}
where $\epsilon$ is a small number. It is tempting to link this threshold to the scheme errors. In some cases however, the process seems to converge but finally oscillates without getting stable, which could lead to false solutions. This is observed for instance close to the mass shedding limit. In practical, we therefore let the field $\deh$ fall down at its minimum, which is generally close to the computer precision. This does not appear detrimental.

 Once the solution (i.e. $\hrho$ and the three constants $C_1$, $C_2$ and $C_3$) is known, several output quantities can be infered, in particular (in dimensionless form):
\begin{itemize}
\item the area of the meridional section $\hat{S} = \iint_{\cal S}{d\ha d\hz}$,
\item the fluid volume $\hat{V} = 2 \pi \iint_{\cal S}{\ha d\ha d\hz}$,
\item the mass $\hat{M} = 2 \pi \iint_{\cal S}{\hrho \ha d\ha d\hz}$,
\item the mean mass density $\langle \hrho \rangle = \frac{\hm}{\hat{V}}$,
\item the angular momentum $\hjcin = 2 \pi \iint_{\cal S}{\hrho \ho \ha^3 d\ha d\hz}$,
\item $\hw$, $\hpi$, $\hcin$, and the Virial parameter $VP$.
\end{itemize}

\subsection{An example with $n \ge 1$}
\label{subsec:example1}

As a first example, we consider a rigidly rotating torus with parameters $n=1.5$, $\hr_A=0.5$, $\hr_B=1$. This case is for instance reported in \cite{hachisu86} and others. For $\Omega=\Omega_0$, the centrifugal potential is quadratic with the radius, namely
\begin{equation}
\begin{cases}
\ho=1,\\
\hphi=-\frac{1}{2}\hr^2.
\end{cases}
\end{equation}
Regarding the numerical grid, we take $\ell=7$ (i.e. $N=128$) and $N_0=2$ which implies for the radial sampling $\hr_2=\hr_A$ and $\hr_{N-2}=\hr_B$. According to Sect. \ref{subsec:grid}, we have
\begin{equation}
\begin{cases}
\hat{h} = 0.00403226\dots\\
\hr_0 =  0.491935\dots \\
\hr_N = 1.008065\dots\\
\hbz_N=\frac{1}{2}(\hr_N-\hr_0) = +0.258065\dots\\ 
\end{cases}
\end{equation}

\begin{figure}
\includegraphics[width=8.7cm,bb=76 275 561 526,clip=]{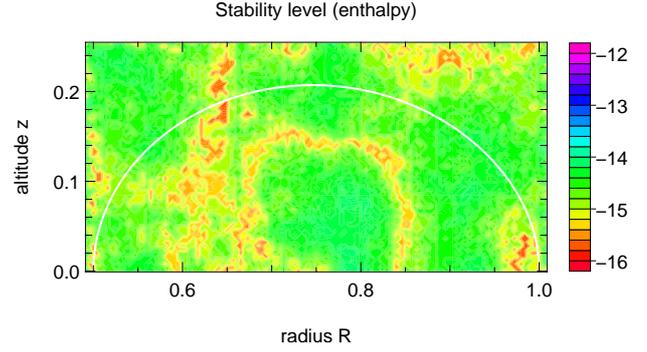}
\caption{Deviation $\deh$ at convergence for the torus considered in Sec. \ref{subsec:example1}, i.e. $n=1.5$, $\hr_A=0.5$, $\hr_B=1$, $\ell=7$ and $N_0=2$.}
\label{fig:ehent_01.ps}
\end{figure}

\begin{figure}
\includegraphics[width=8.7cm,bb=76 300 561 526,clip=]{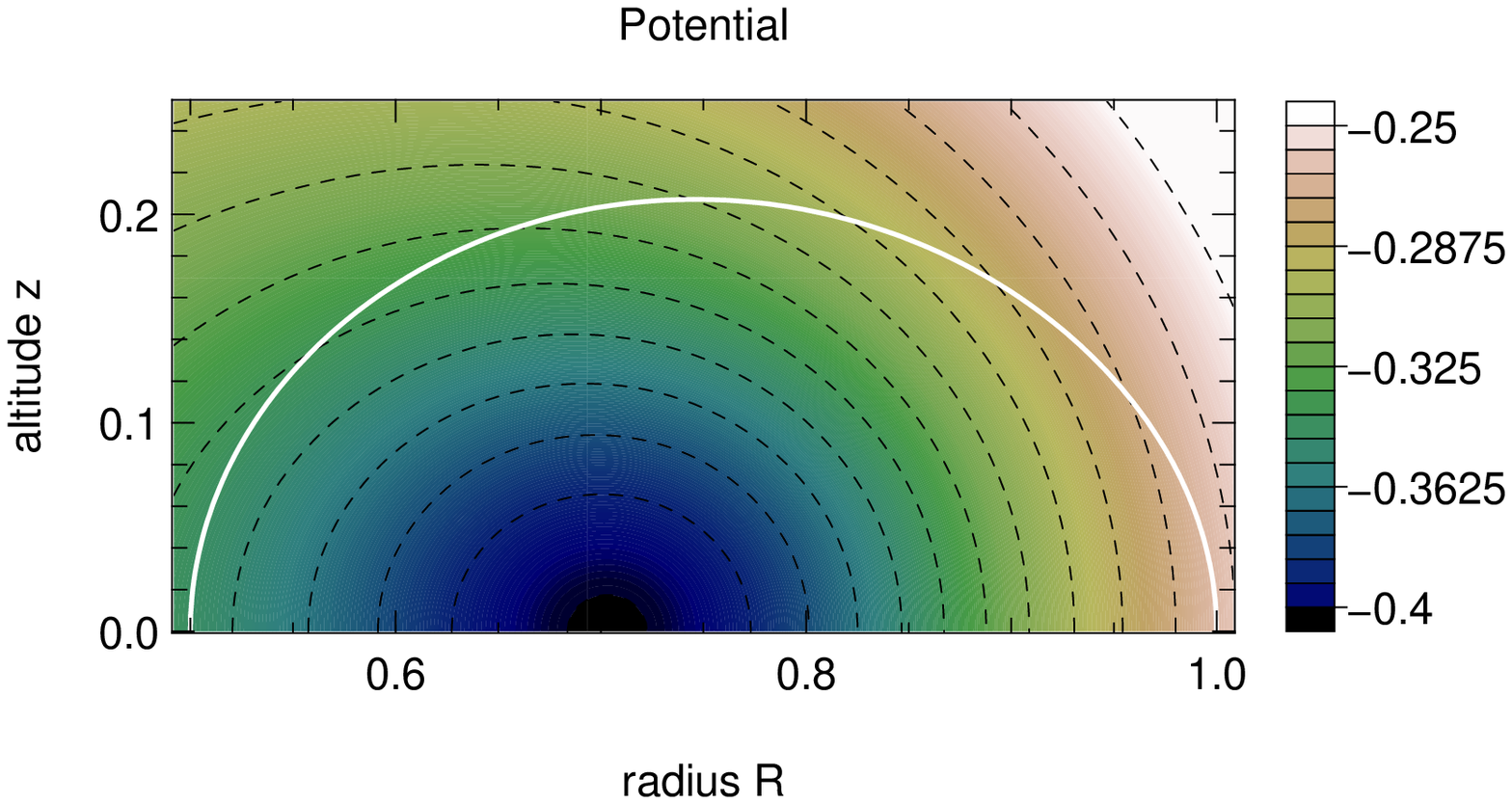}\\
\includegraphics[width=8.7cm,bb=76 300 561 526,clip=]{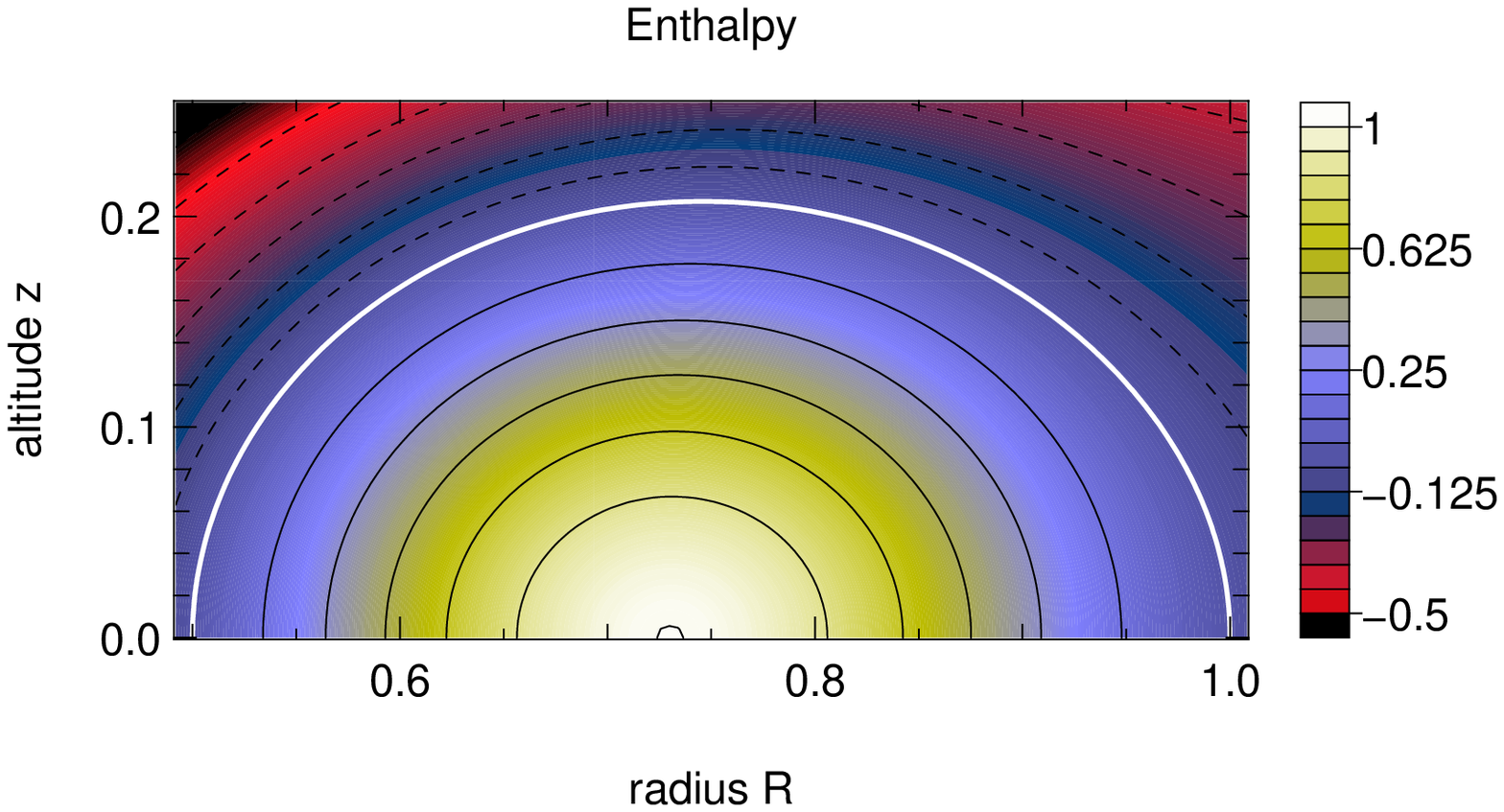}\\
\includegraphics[width=8.7cm,bb=76 275 561 526,clip=]{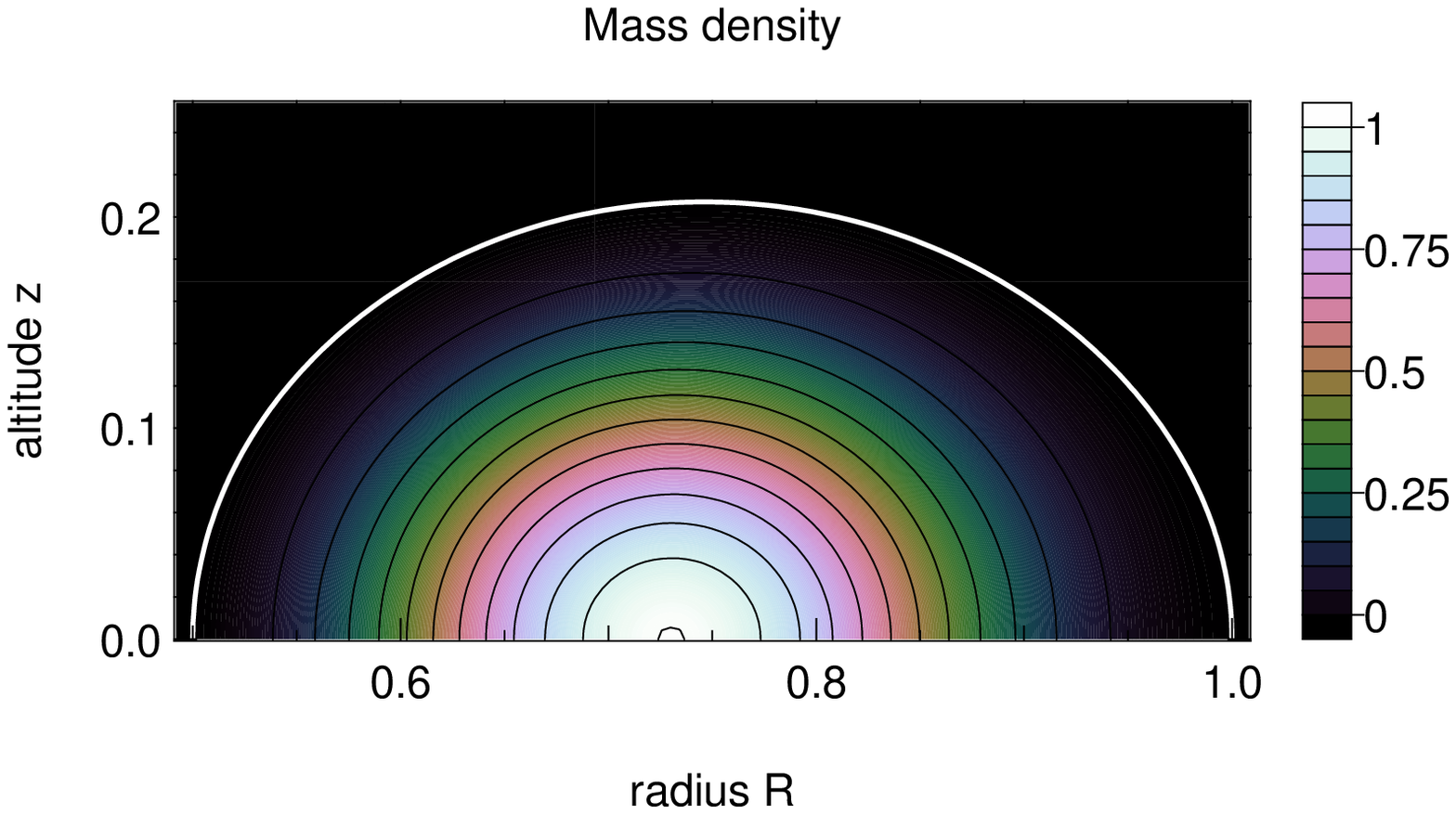}
\caption{Same legend as for Fig. \ref{fig:ehent_01.ps} but for the potential $\hpsi$, the enthalpy $\hh$ and the mass density $\hrho$ at equilibrium ({\it from top to bottom}).}
\label{fig:config_ref.eps}
\end{figure}

\begin{figure}
\includegraphics[width=8.5cm,bb=21 46 709 522,clip=]{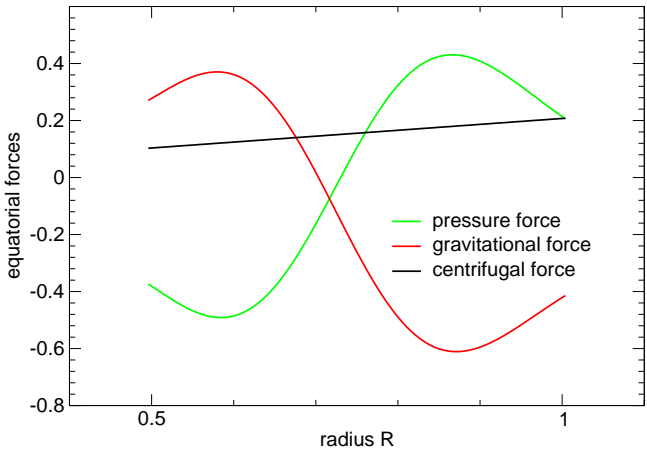}
\caption{Forces at the equatorial plane versus the radius for the torus considered in Fig. \ref{fig:config_ref.eps}.}
\label{fig:eqforces}
\end{figure}

\begin{figure}
\includegraphics[width=8.7cm,bb=76 300 561 526,clip=]{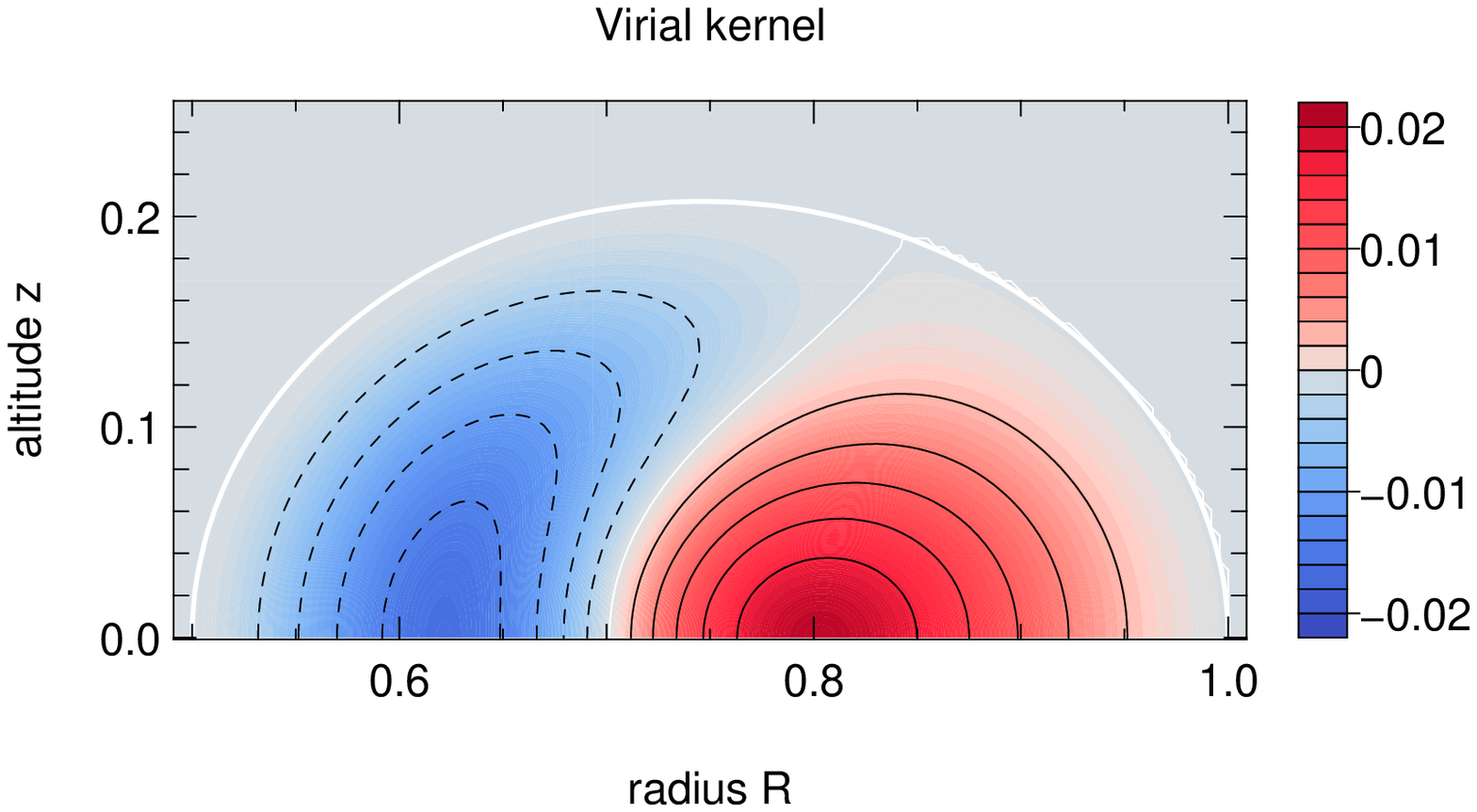}
\includegraphics[width=8.7cm,bb=76 275 561 526,clip=]{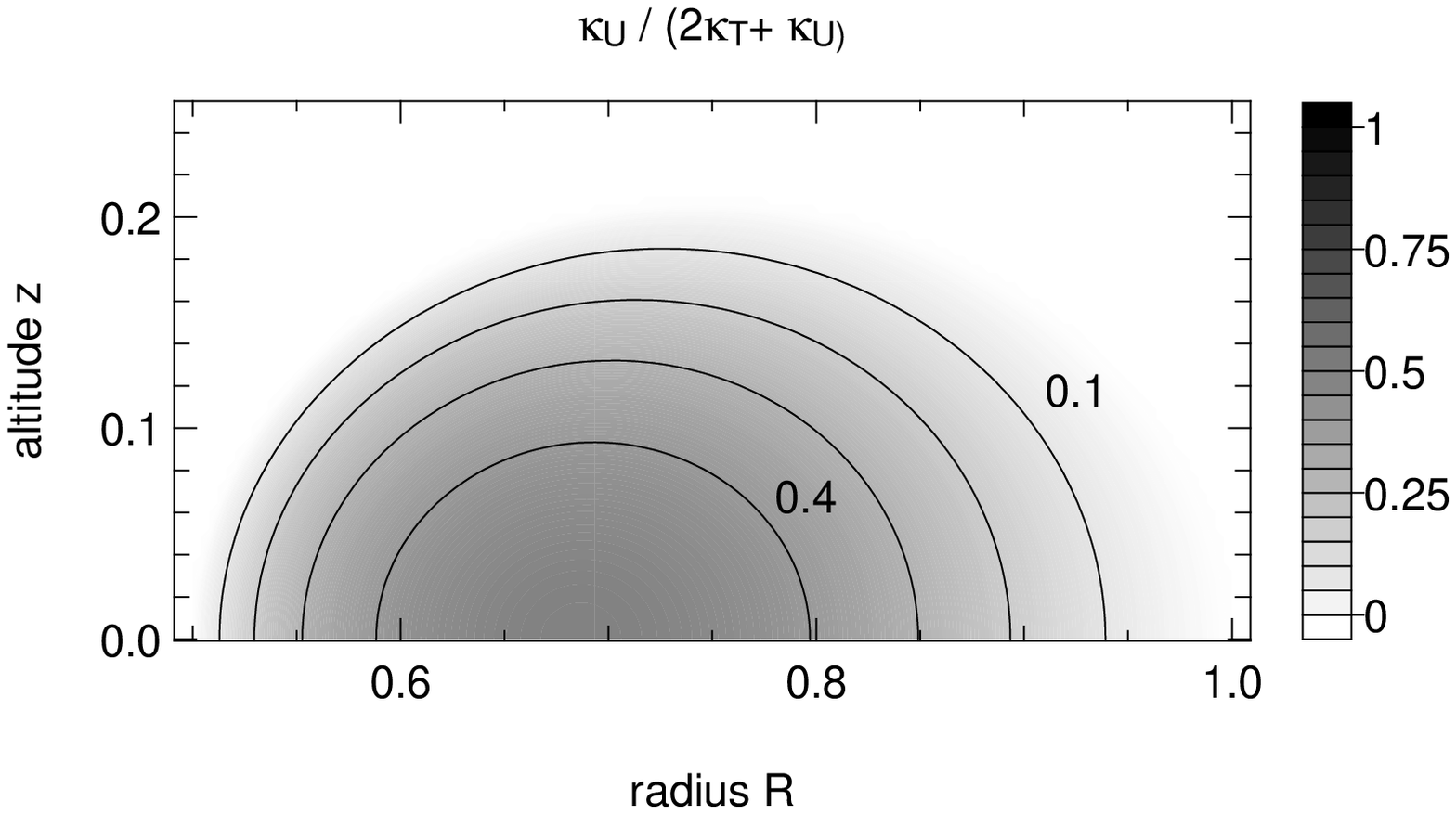}
\caption{Same legend as for Fig. \ref{fig:config_ref.eps} but for the virial kernel $\kv$ ({\it top}) and the ratio $\frac{\kappa_U}{2\kappa_T+\kappa_U}$ ({\it bottom}).}
\label{fig:kernels_ref.eps}
\end{figure}

For these parameters, the absolute error in potential values is expected to be of the order of $\sim 2 \times 10^{-5}$ (see Sect. \ref{subsec:secondorderpoisson}), which is transmitted to $\hh$, $\hrho$, constants $C_1$, $C_3$, $C_3$, $VP$, etc. This means $4$ correct digits typically, which is typical at such a spatial resolution \citep{eri78,hachisu86,ye06}. The equilibrium solution is found after $65$ iterations.  The remarkable capability of convergence of the SCF-loop is shown in Fig. \ref{fig:ehent_01.ps} which displays $\deh$ a few steps after. Figure \ref{fig:config_ref.eps} displays $\hpsi$, $\hh$ and $\hrho$ at convergence. We notice the slight asymetry of the enthalpy and density maps (point M is closer to edge A than edge B), due to curvature effects. Centrifugation shifts the gravitational potential well towards the inner edge. Equatorial forces are plotted in Fig. \ref{fig:eqforces}. We see that the zero-gravity and zero-pressure points do not coincide. The Virial kernel and the quantity $\frac{\kappa_U}{2\kappa_T+\kappa_U}$ are displayed in Fig. \ref{fig:kernels_ref.eps}. The gravitational energy density is more concentrated at the inner edge of the torus, while the internal and centrifugal energies dominate in the outer region. The kinetic term always exceeds the pressure term, and this is especially pronounced close to the fluid boundary. Table \ref{tab:cpredtohachisu} shows that output quantities are in agreement with \cite{hachisu86} who works with Simpson's $4$th-order quadrature rule in spherical coordinates and rather low covering factor. Successful comparisons are observed for many other configurations. Deviations appear close to critical rotation and in the thin ring limit as well.

\begin{table}
\centering
\begin{tabular}{lll}\\
  quantity                  & Hachisu (1986)     & this work     \\ \hline
  covering factor $\Lambda$ & $0.091^*$            & $0.608$       \\
  $C_1$                     & $0.0842$             & $0.08422$     \\
  $C_2$ (i.e. $\Omega_0^2)$  & $0.207$              & $0.2068$      \\
  $-C_3$                     &                      & $0.3691$       \\
  $\hr_e$                    &  $0.5$              & $0.5$ \\
  $\hr_p$                    &                     & $0.2071$ \\
 $\hbsur$                   &                     & $0.1619$ \\
  $\hv$                      & $0.762$             & $0.7615$  \\ 
  $\hm$                      & $0.219$             & $0.2188$  \\
  $\langle \hrho \rangle$    &  $0.287^*$           & $0.2874$ \\
  max. enthalpy              & $0.0842$             & $0.08422$ \\
  max. pressure              & $0.0337$             & $0.03369$ \\
  max. density               &  $0.0245^*$          & $0.02444$ \\
  $\sqrt{C_2}\hjcin$         & $0.0562$             & $0.05613$  \\
  $C_2 \hcin$                & $0.0128$             & $0.01276$  \\
  $-\hw$                     & $0.0401$             & $0.04005$ \\
  $\frac{C_1}{n+1}\hu$                      & $0.0145$             & $0.01453$  \\
  $\log(VP)$                 & (?)                  & $-4.42$   \\
 iterations & ?        & $65$   \\\hline
\end{tabular}
\caption{Comparison with Hachisu (1986). At the actual resolution, only four figures are guaranteed ($^*$estimated).}
\label{tab:cpredtohachisu}
\end{table}

\subsection{$VP$ vs. covering factor}

By varying $N_0$, we modify the covering factor (fluid section-to-grid area ratio) defined by
\begin{equation}
\frac{\hat{S}}{(\hr_N-\hr_0)(\hbz_N-\hbz_0)} \equiv \Lambda.
\end{equation}

\begin{figure}
\includegraphics[height=8.7cm,angle=-90,bb=76 20 570 721,clip=]{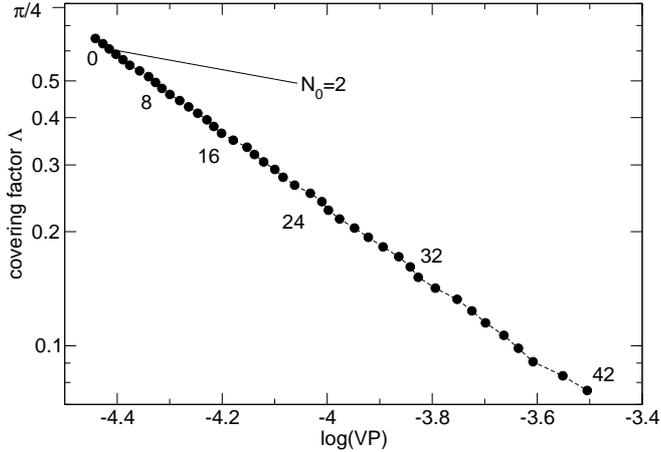}
\caption{Variation of the Virial Parameter with the covering factor $\Lambda$. Values of $N_0$ are reported (see Fig. \ref{fig:config_ref.eps} for $N_0=2$ and Fig. \ref{fig:h_ref2.ps} for $N_0=42$)}
\label{fig:vtvsn0.ps}
\end{figure}

\begin{figure}
\includegraphics[width=8.7cm,bb=76 300 561 526,clip=]{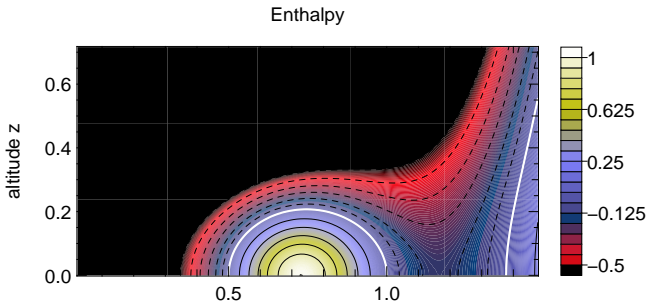}\\
\caption{Same legend as for Fig. \ref{fig:config_ref.eps}b but with $N_0=42$.}
\label{fig:h_ref2.ps}
\end{figure}

In the above example, $\Lambda \approx 0.61$. When increasing $N_0$ (from $0$ to $N'/3$), the resolution of the fluid is decreased, which decreases $\Lambda$. Figure \ref{fig:vtvsn0.ps} displays $VP$ vs. $\Lambda$ in log. scale. While the sensitivity remains weak, there is a net linear correlation, and we find $\log VP \approx -4.58 - 0.99 \log \Lambda$. The more resolved the fluid, the better the Virial parameter. Figure \ref{fig:h_ref2.ps} shows the enthalpy at equilibrium for the largest value $N_0=42$. 

\subsection{$VP$ as a reliable indicator for precision}

\begin{figure}
\includegraphics[height=8.7cm,angle=-90,bb=55 0 583 721,clip=]{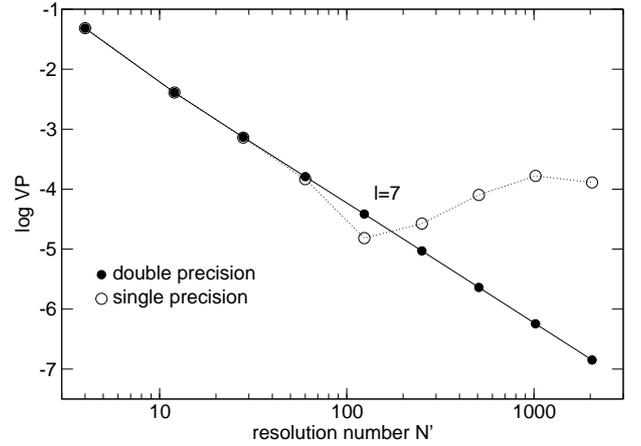}
\caption{Variation of the Virial parameter $VP$ with the resolution number $N'$ for $\ell \in [3,11]$.}
\label{fig:vtvsl.ps}
\end{figure}

We have considered various numerical resolutions $\hat{h}$ for the finest grid, which is obtained by varying $\ell$, according to Eq.(\ref{eq:gres}). We have explored the range $\ell \in [3,11]$. The convergence of the SCF-loop is almost unchanged, with $65$ iterations, whatever $\ell$. The Virial parameter is plotted versus $N'$ in Fig. \ref{fig:vtvsl.ps} for both single and double precision. The correlation is remarkable. A linear regression yields $\log VP \approx -0.16915 - 2.02367 \log N'$, or
\begin{equation}
VP \approx 0.339 \times \hat{h}^{2.02}
\end{equation}
which means that the Virial parameter is $2$nd-order accurate. For single precision, accuracy is maximum for $\ell \sim 7$ and deteriorates for finer grids, in agreement with the error analysis (see Sect. \ref{subsec:secondorderpoisson}).

\begin{figure}
\includegraphics[width=8.5cm,bb=39 46 706 522,clip=]{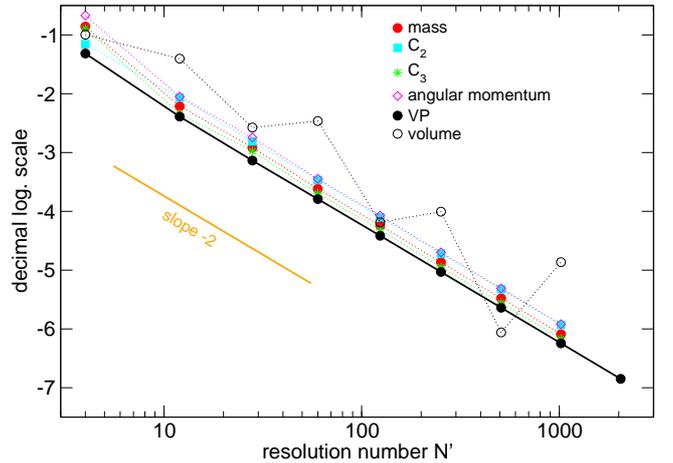}
\caption{Precision observed on a few output quantities versus the resolution number $N'$. The volume is not precsely determined unless the fluid boundary is accounted for (see Sect. \ref{sec:subgrid}).}
\label{fig:C2vsl.eps}
\end{figure}

We can easily check that all output quantities are improved when the resolution is increased by comparing the results at $\ell-1$ and $\ell$, namely 
\begin{equation}
 \left| 1-\frac{\hf(\ell-1)}{\hf(\ell)} \right| \sim \left(\frac{d \ln |\hf|}{d \ell} \right)_{\ell-1}.
\end{equation}
Figure \ref{fig:C2vsl.eps} shows this quantity for the mass, the two constants $C_2$ and $C_3$, the angular momentum $\hjcin$, and the volume $\hv$. We see that the Virial parameter is a good tracer of precision for most output quantities, within a factor less than $2$ typically. The exception is for the volume, and for the meridional section area as well (not shown). This is in agreement with the discussion in Sec. \ref{subsec:quad}.

\begin{figure}
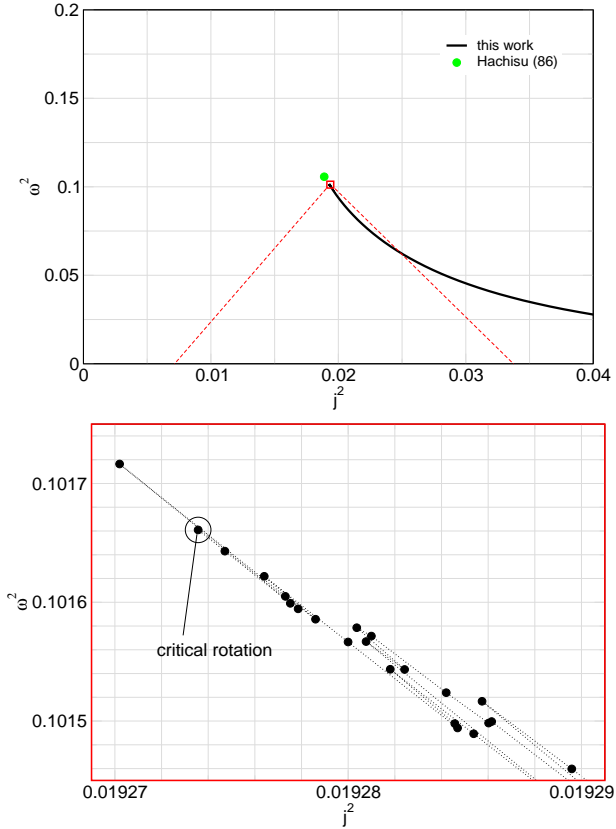

\includegraphics[width=8.cm,bb=0 25 730 531,clip=]{hachisu.eps}\\
\includegraphics[width=8.cm,bb=-9 34 716 523,clip=]{hachisub.eps}
\caption{The $\omega^2-j^2$ diagram for $n=1.5$ at two different scales. Near critical rotation ({\it bottom panel}), the sequence is not well defined unless the fluid boundary is detected and accounted for. The value given in Hachisu (1986) stands outside the plot.}
\label{fig:hachisub}
\end{figure}

\subsection{Uncertainty of the $\omega^2-j^2$ diagram near critical rotation}

A classical manner to visualize and compare equilibria obtained for different triplets $(C_1,C_2,C_3)$ is the $\omega^2-j^2$ diagram \citep{chandra73,hachisu86}, where
\begin{equation}
\begin{cases}
j^2 =\frac{1}{4\pi G \langle \rho \rangle}\frac{J^2}{M^2 V^{4/3}},\\
\omega^2 = \frac{1}{4\pi G \langle \rho \rangle}\Omega_0^2,
\end{cases}
\end{equation}
and $J = \rho_0 \Omega_0 L^5 \hjcin$. Using dimensionless quantities, we have
\begin{equation}
(j^2,\omega^2) = \frac{C_2}{4\pi \hm} \times \left(\frac{\hjcin^2}{\hm^2 \hv^{1/3}},\hv\right).
\label{eq:omega2j2}
\end{equation}

For one-ring sequences obtained by varying the torus axis ratio (while $n$ is held fixed), $\omega$ typically decreases when $j$ increases \citep{es81,hachisu86,ansorg03}. Figure \ref{fig:hachisub} displays the diagram obtained by decreasing $\hr_A/\hr_B$. We notice that the ``trajectory'' is hazardeous close to the sequence end. This effect is a direct consequence of the uncertainty in the volume. Critical rotation, where the pressure gradient vanishes at point B (i.e. gravity and centrifugation are just balanced), is found for $\hr_A/\hr_B \sim 0.3238$, against $0.325$ in \cite{hachisu86}.

\begin{table}
\centering
\begin{tabular}{lrl}\\
                  &                       \\
  quantity        & Petroff \& Horatschek (08)     & this work   \\ \hline
  covering factor & $1$                 & $0.731$  \\
  $C_1$             &                   & $0.005363$  \\
  $C_2$ (i.e. $\Omega_0^2)$ & $0.01499$ & $0.01499$ \\
  $-C_3$            &                   & $0.04055$   \\
  $i_M$            &                    & $63$        \\
  $\hh_M$           &                   & $1$         \\
  $\hbsur$        &                     & $0.007786$ \\
  $\hv$             &                    & $0.04647$   \\ 
  $\hm$             &                     & $0.002004$ \\
  $\langle \hrho \rangle$ &               & $0.4312$ \\
  max. enthalpy          &                & $0.005363$ \\
  max. pressure         &                 & $0.002682$ \\
  max. density         &                  & $0.005363$  \\
  $\sqrt{C_2}\hjcin$        &             & $0.002216$ \\
  $C_2 \hcin$         &                   & $0.0001357$ \\
  $-\hw$           &                      & $0.0003720$ \\
  $\frac{C_1}{n+1}\hu$           &        & $0.0001006$  \\
  $\log(VP)$     &    & $-4.75$   \\
 iterations &       & $28$    \\\\
  $\hm \left(\frac{n+1}{C_1}\right)^{3/2}$ & $144.3$ & $144.3$ \\
 $\hjcin \left(\frac{n+1}{C_1}\right)^{5/2}$        & $5953$     & $5952$ \\
  $\hcin \left(\frac{n+1}{C_1}\right)^{5/2}$         & $364.5$    & $364.4$ \\
  $-\hw \left(\frac{n+1}{C_1}\right)^{5/2}$          & $999.2$    & $999.0$  \\
  $\hpi \left(\frac{n+1}{C_1}\right)^{5/2}$           & $90.07$   & $90.06$  \\\hline
\end{tabular}
\caption{Results obtained for $n=1$ and $\hr_A/\hr_B=0.9$ compared with the spectral approach by Petroff \& Horatschek  (2008). Same grid parameters as for Tab. \ref{tab:cpredtohachisu}}
\label{tab:cpredtopetroff}
\end{table}

\subsection{A case with $n=1$}
\label{subsec:example2}

\begin{figure}
\includegraphics[width=8.7cm,bb=-6 24 744 530,clip=]{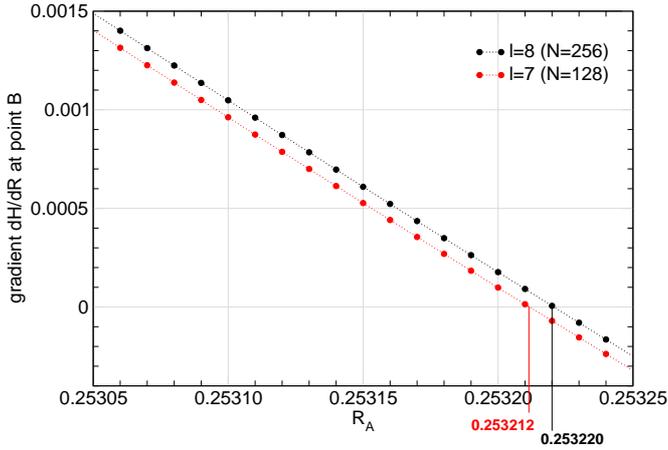}
\caption{Enthalpy gradient at point B near critical rotation for $n=1$ and two numerical resolutions.}
\label{fig:gradhn1.eps}
\end{figure}

\begin{figure}
\includegraphics[width=8.7cm,bb=76 300 561 526,clip=]{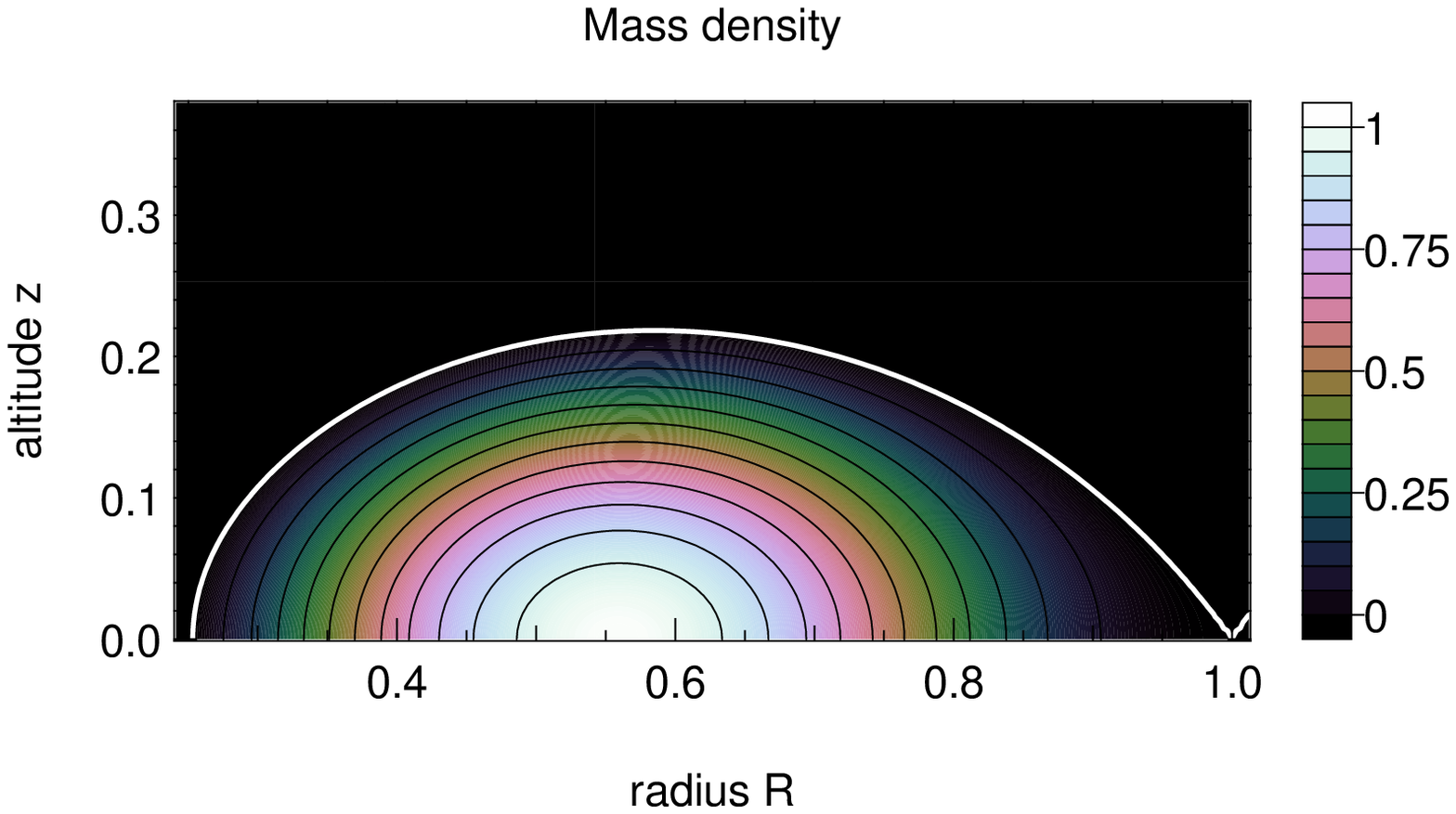}
\includegraphics[width=8.7cm,bb=76 300 561 526,clip=]{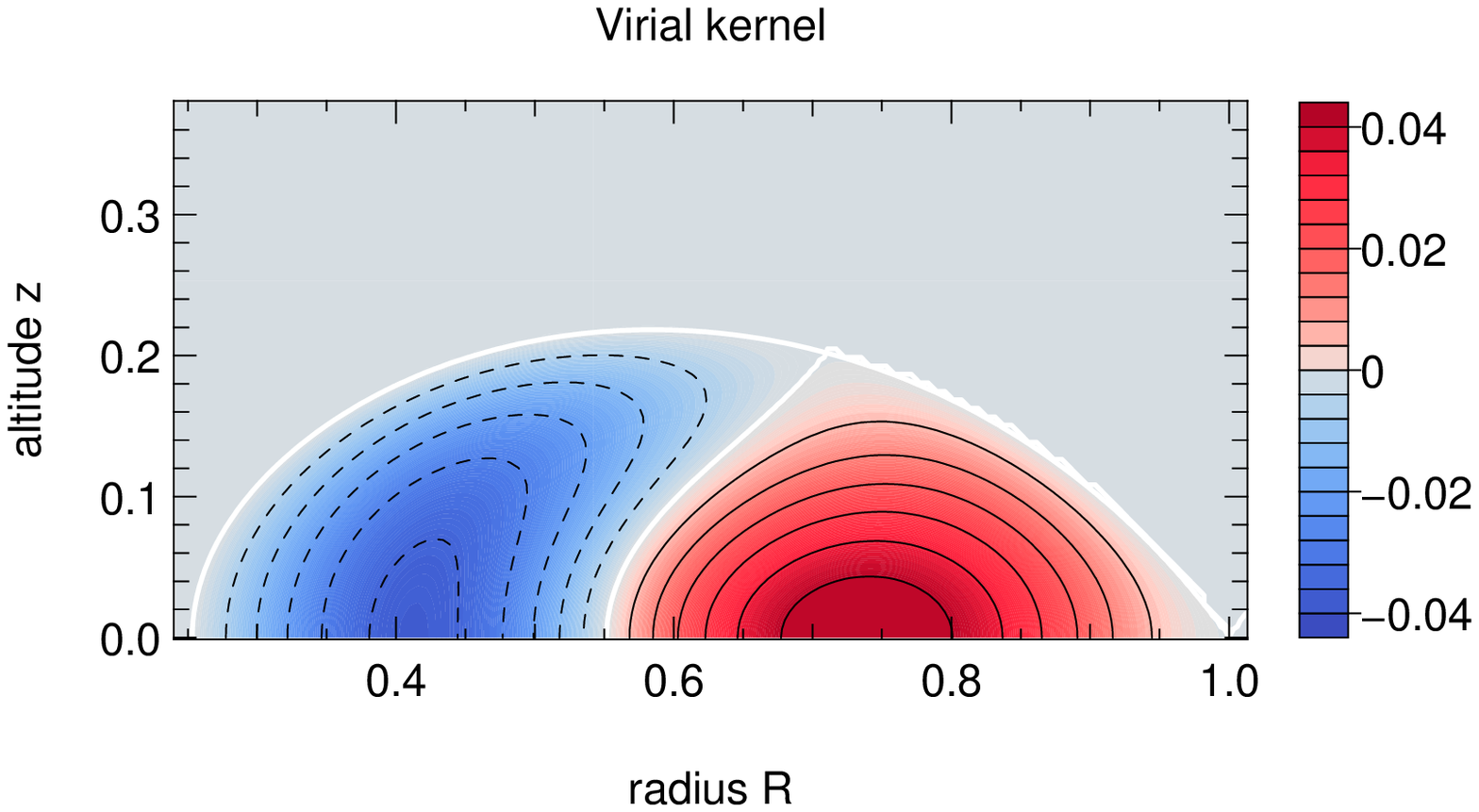}
\includegraphics[width=8.7cm,bb=76 275 561 526,clip=]{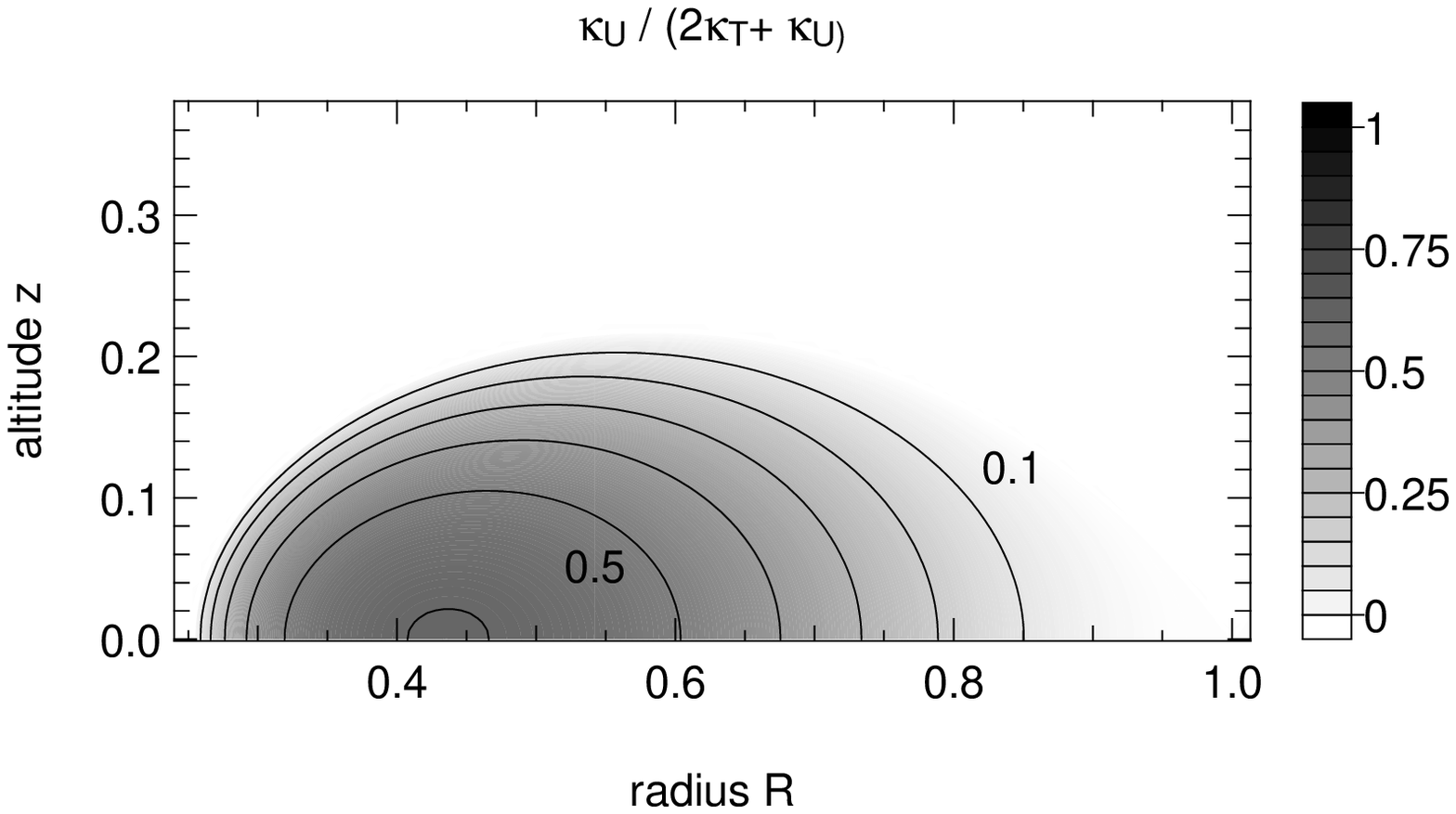}
\caption{Same legend as for Figs. \ref{fig:config_ref.eps} and \ref{fig:kernels_ref.eps} but for $n=1$ at critical rotation where $\hr_A/\hr_B \approx 0.253221$.}
\label{fig:config_ref3.eps}
\end{figure}

As a second example, we consider a rigidly rotating torus with $n=1$ and $\hr_A/\hr_B=0.9$. The solution is found after $28$ SCF-iterations. The internal structure resembles very much the one presented above. The torus boundary is much closer to a circle, with a polar radius $\hr_p=0.0495$. Output quantities are given in Tab. \ref{tab:cpredtopetroff}. The agreement with the results obtained from a spectral method by \cite{petroff08} is remarkable. From this equilibrium, we can reach the critical rotation by decreasing the radius of the inner edge (the number of SCF-iterations generally increases). Figure \ref{fig:gradhn1.eps} shows $\nabla_R \hh$ at the outer edge versus $\hr_A$. There is obviously a slight resolution effect, but the gradient vanishes for $\hr_A/\hr_B \sim 0.2532$ which is also the solution of \cite{petroff08} obtained from a spectral approach. Figure \ref{fig:config_ref3.eps} displays the mass density, Virial kernel and fraction $\frac{\kappa_U}{2\kappa_T+\kappa_U}$. The fluid core is dominated by the gravitational energy, and it remains more marked close to the inner edge. We see the elongated structure at the outer edge where matter is in pure Keplerian rotation.

\begin{figure}
\includegraphics[width=8.7cm,bb=76 300 561 526,clip=]{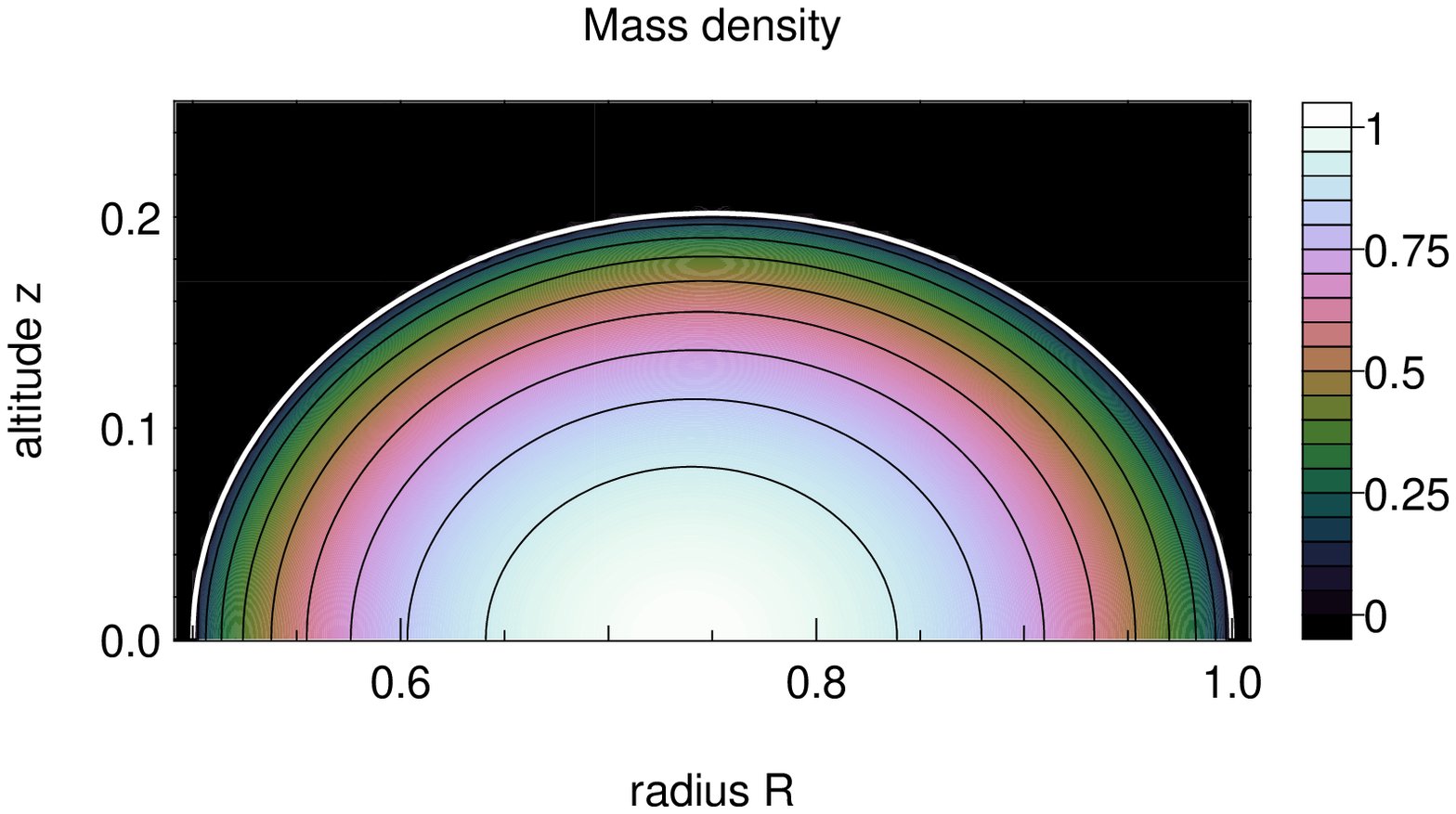}
\includegraphics[width=8.7cm,bb=76 300 561 526,clip=]{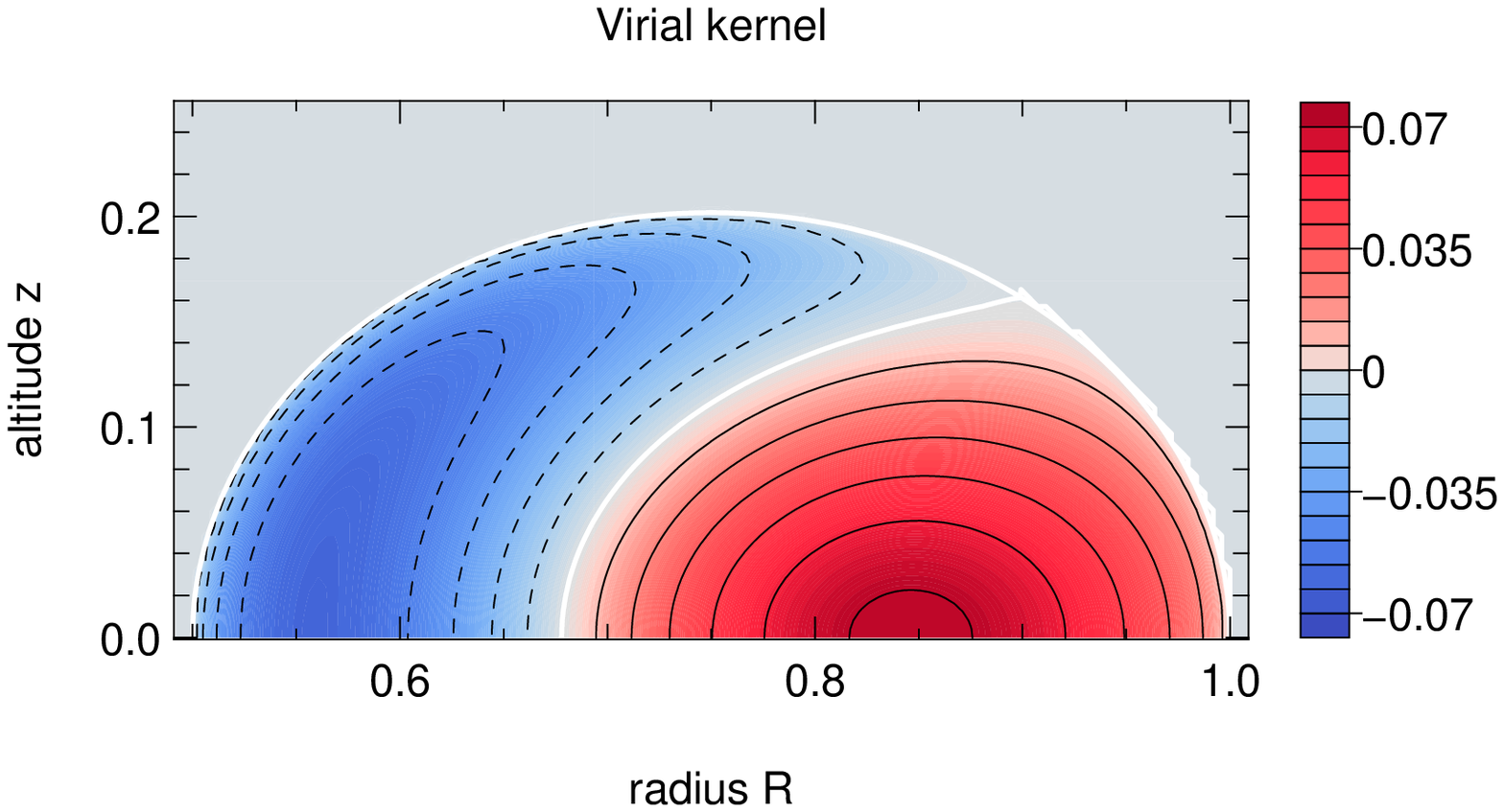}
\includegraphics[width=8.7cm,bb=76 275 561 526,clip=]{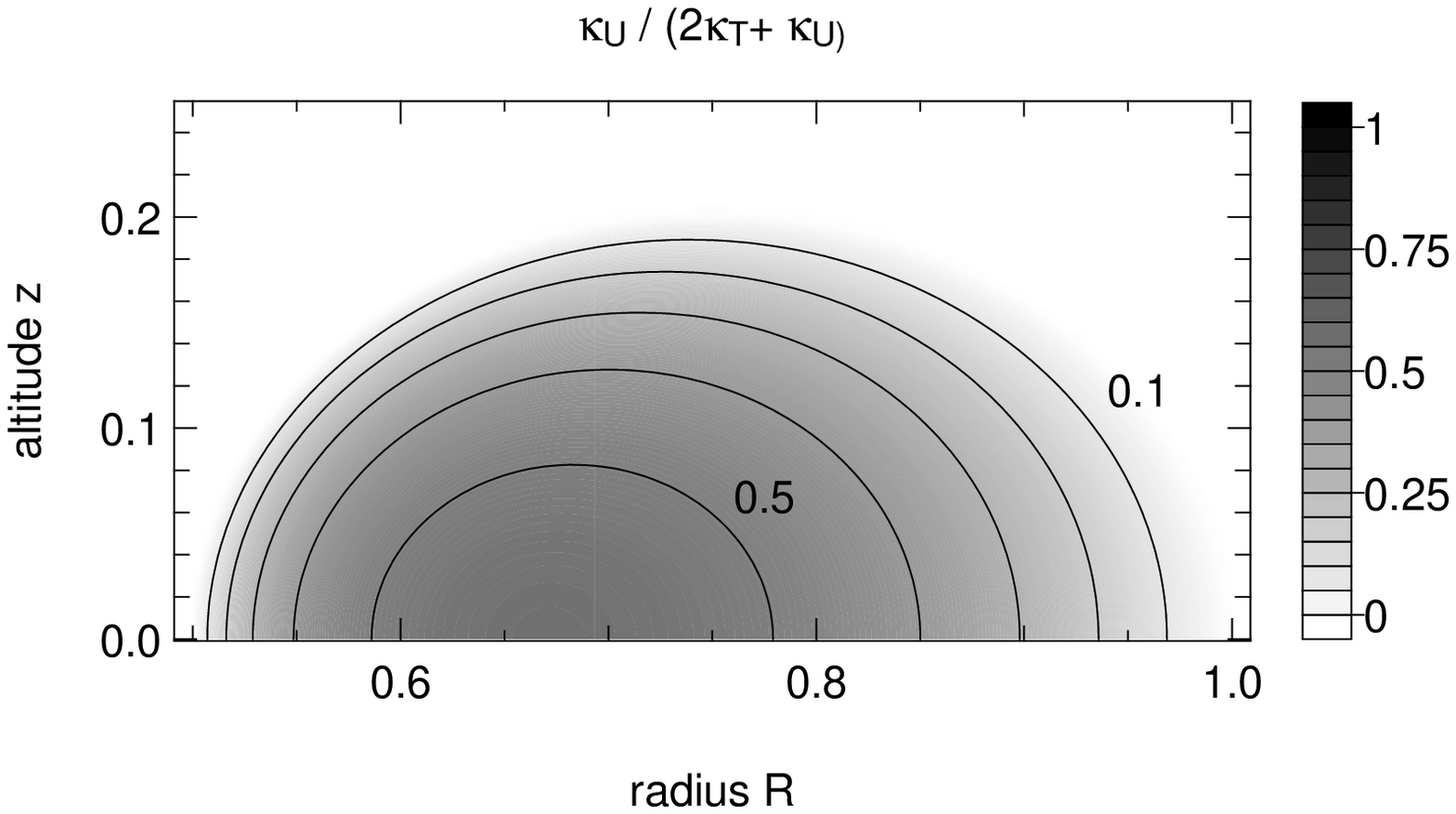}
\caption{Same legend as for Fig. \ref{fig:config_ref.eps} but for $n=0.5$.}
\label{fig:rho_reflowerthanone.eps}
\end{figure}

\subsection{Case with a hard EOS}
\label{subsec:example3}

\begin{figure}
\includegraphics[width=8.5cm,bb=39 46 706 522,clip=]{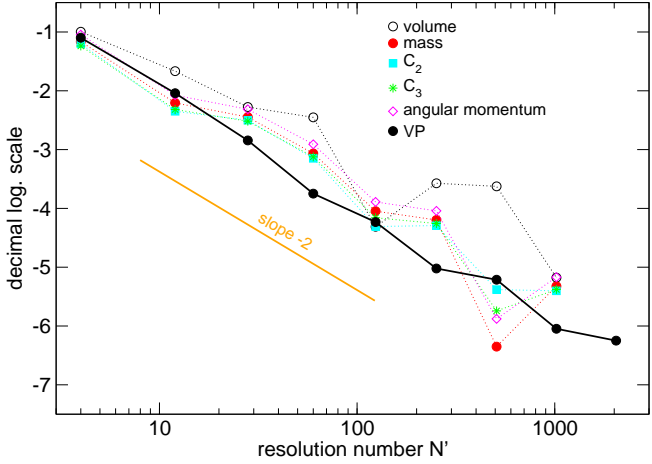}
\caption{Same legend as for Fig. \ref{fig:C2vsl.eps} but for $n=0.5$.}
\label{fig:C2vsl_n05.eps}
\end{figure}

We consider the same parameters as in Sect. \ref{subsec:example1} but with $n=0.5$. This is a typical case where $\rho$ has an infinite gradient onto $\Gamma$. The equilibrium solution is now found after $34$ iterations. In general, convergence is faster with low indices $n$. The mass density, Virial kernel and ratio  $\frac{\kappa_U}{2\kappa_T+\kappa_U}$ are displayed in Fig. \ref{fig:rho_reflowerthanone.eps}. Again, the comparison with \cite{hachisu86} does not reveal any deviations for the first three digits. The Virial test indicates $\log(VP)\approx-4.2$.  This is astonishing since, as argued in Sec. \ref{subsec:quad}, quadratures are expected to be inaccurate for such polytropic index. This illustrates the fact that the Virial test does not prove that the actual solution is the good one, but that it is numerically clean and self-consistent. We have run the code for various resolution numbers where $\ell \in [3,11]$ and computed the variation of some quantities, as done above. The results are displayed in Fig. \ref{fig:C2vsl_n05.eps} and are to be compared with Fig. \ref{fig:C2vsl.eps}. While the global trend is still there, the order is oscillating. All quantities are concerned, not only geometrical ones.

\section{Subgrid approach: method and examples}
\label{sec:subgrid}

\subsection{Detection of the fluid boundary $\Gamma$. Case of a soft EOS}

Detecting the fluid boundary is basically achieved by scanning the whole grid radially or/and vertically and searching for nodes where the sign of enthalpy changes from one node to the other. There are degenerescences if $\Gamma$ is not connex, which case is met for instance when matter is pinched at one or more places around the equatorial plane \citep{ansorg03}. We then use a more straightforward method based on the Freeman chain code \citep[e.g.][]{freeman61onthe} : $\Gamma$ is gradually revealed by visiting neighbouring cells following the same searching sequence. It is in fact much faster than by direct scan because the grid is not explored as a whole. The algorithm starts at point A and fastly reaches point B, which generates $N_\Gamma$ points with coordinates $(\hrg,\hzg)$ where $\hh=0$ (the root finding method is second-order accurate here). If $\Gamma$ has length $\hpe$, then $N_\Gamma \sim \hpe/\hat{h}$ typically.

\begin{figure}
\includegraphics[width=8.7cm,bb=0 0 478 337,clip=]{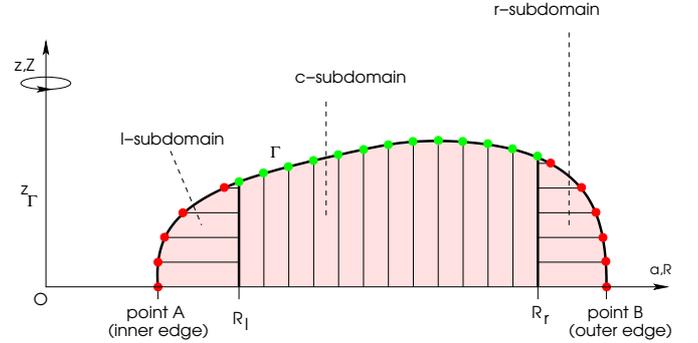}
\caption{Double integrals over the fluid section are improved by dividing the fluid into three sub-domains. Close to the inner and outer edges, the radial quadrature is performed first. In the c-subdomain, the vertical quadrature is performed first.}
\label{fig:tore_et_ell_subdomains.eps}
\end{figure}

As a matter of fact, we do not need all the $\Gamma$-nodes so detected to improve the quality of double integrals. Basically, the inner and outer edges where $d\hzg/d\hr$ is large must be oversampled. A good compromise consists in dividing the fluid section into $3$ subdomains:
\begin{itemize}
\item a ``l-subdomain'' for the inner edge,
\item a ``r-subdomain'' for the outer edge,
\item a ``c-subdomain'' located in between,
\end{itemize}
which requires two intermediate radii $\hr_l$ and $\hr_r$, as depicted in Fig. \ref{fig:tore_et_ell_subdomains.eps}. For the l- and r-subdomains, we integrate in the radial direction first by considering the boundary nodes, then in the vertical direction. For the c-subdomain, we proceed in the reverse way. The double integral is therefore written in the form
\begin{flalign}
\nonumber
\frac{1}{2}\iint_{\rm fluid}{\hf d \ha d\hz}&=\underbrace{\int_0^{\hzg}{ d\hz \int_{\hrg(\hz)}^{\hrl(\hz)}{\hf d \ha}}}_{\text{l-subdomain}}+ \underbrace{\int_\hrl^\hrr{d \ha \int_0^{\hzg(\ha)}{\hf  d\hz }}}_{\text{c-subdomain}}\\
& \qquad + \underbrace{\int_0^{\hzg}{d\hz \int_\hrr^{\hrg(\hz)}{\hf d \ha}}}_{\text{r-subdomain}},
\label{eq:quadwithgamma}
\end{flalign}
and executed as is (the factor $2$ accounts for matter below the equatorial plane). The quadrature scheme is almost the same as in Eq.(\ref{eq:trap}), except at the fluid boundary where two nodes have a spacing smaller than $\hat{h}$. This technique is quite easy to implement. The two radii $\hr_l$ and $\hr_r$ could change inside the SCF-iterations (depending on $d\hzg/d\hr$). For the present study, however, we simply take  
\begin{flalign}
\begin{cases}
\hr_l = \hr_A + \frac{1}{2} \hr_e,\\
\hr_r = \hr_B - \frac{1}{2} \hr_e.
\end{cases}
\end{flalign}

\begin{figure}
\includegraphics[width=8.cm,bb=28 46 707 522,clip=]{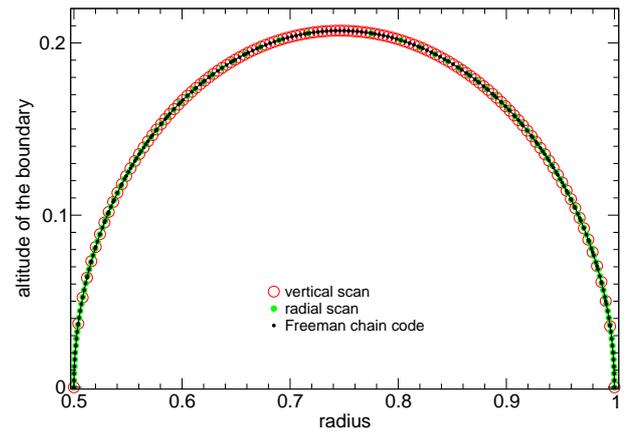}
\caption{Fluid boundary for the torus with soft EOS considered in Sect. \ref{subsec:example1}. By performing a vertical scan, the precise curvature of $\Gamma$ close to the inner and outer edges is not accessible. Similarily, the radial scan does not give much detail of the boundary at the top/bottom of the fluid.}
\label{fig:zplus.eps}
\end{figure}

\begin{figure}
\includegraphics[width=8.5cm,bb=39 46 706 522,clip=]{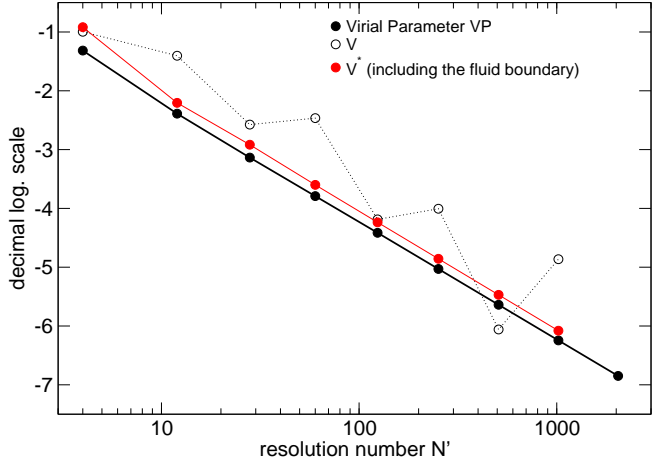}
\caption{Same legend as for Fig. \ref{fig:C2vsl.eps} but when the fluid volume is computed at $2$nd-order by accounting for the boundary ({\it red}).}
\label{fig:C2vsl_wcont.eps}
\end{figure}

\begin{figure}
\includegraphics[width=8.cm,bb=-9 34 716 523,clip=]{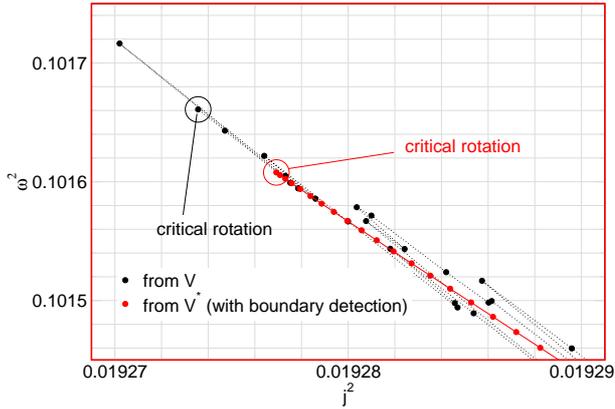}
\caption{Same conditions as for Fig. \ref{fig:hachisub} but when the the fluid boundary is detected and accounting for.}
\label{fig:hachisuc.eps}
\end{figure}

\subsection{Results for $n=1.5$}

We go back to the example considered in Sect. \ref{subsec:example1} and recompute the solution including the detection of $\Gamma$, for different numerical resolutions. BVs that feed the Poisson-solver are then estimated from Eq.(\ref{eq:quadwithgamma}) at each time step of the SCF-loop. At convergence, output quantities (Viral parameter, fluid volume, etc.) are computed in a similar manner. Figure \ref{fig:zplus.eps} shows the result obtained for the example given in Sect. \ref{subsec:example1}. We find $N_\Gamma=227$ for $\ell=7$. Figure \ref{fig:C2vsl_wcont.eps} shows the new volume, denoted $\hv^*$ versus $N'$. We clearly see that error is now fully $2$nd-order accurate for the volume (and other geometrical quantities). As a consequence, the equilibrium sequence near critical rotation is now much better described as Fig. \ref{fig:hachisuc.eps} shows.

\subsection{Detection of the fluid boundary combined with kernel splitting. Case of a hard EOS}

\begin{figure}
\includegraphics[width=8.2cm,bb=35 10 330 250,clip=]{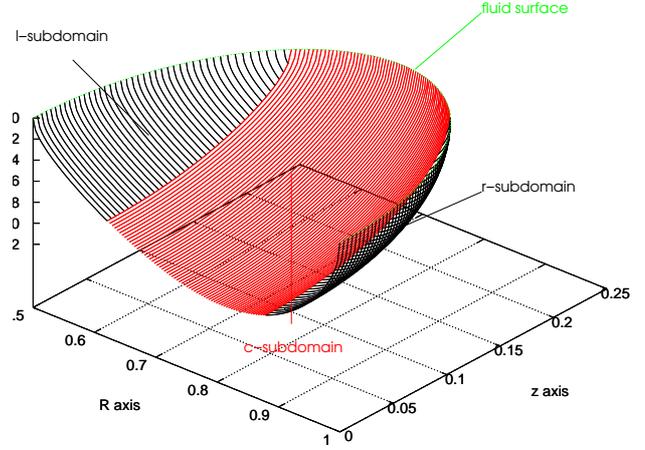}
\caption{Poisson kernel $\kappa \rho$ limited to the half plane $Z \ge 0$ versus $a$ and $z$ for the boundary value at $(\hr_N,0)$ for the torus with hard EOS considered in sect. \ref{subsec:example3}.}
\label{fig:dsmpict.eps}
\end{figure}

If we apply the preceeding recipe to configurations with $n\le 1$, the second-order is still not recovered for the reason evoked in Sect. \ref{subsec:quad}: the presence of infinite gradients all along the fluid boundary, as a direct consequence of the polytropic assumption, makes the numerical integration too uncertain. This is illustrated in Fig. \ref{fig:dsmpict.eps} which displays the full integrand $\kappa \hrho$ in Eq.(\ref{eq:psibcs}) obtained when computing one of the BVs for $n=0.5$ at convergence. Infinite gradient are clearly visible all along the fluid boundary. The problem is not fixed by increasing the order of the quadrature scheme. A convenient approach consists in treating the underlying power-law by analytical means. Actually, by using the generic $q$-coordinate (which is $a$ or $z$ depending on the subdomain; see below), we have
\begin{equation}
\hf(q)\hrho(q)=\hf_0\hrho_0 \left(\frac{q-q_\Gamma}{q_0-q_\Gamma}\right)^n + \delta (\hf \hrho ) 
\end{equation}
where $q_0>q_\Gamma$ is a pivot, $\hf_0\hrho_0=\hf(q_0)\hrho(q_0)$, and $\delta (\hf \hrho)$ is the residue, i.e. the deviation to the power-law. By integrating over $q$, we get
\begin{flalign}
\nonumber
\int_\qg^q{\hf(q') \hrho(q') d q'} &=  \int_\qg^q{\hf_0\hrho_0 \left(\frac{q-q_\Gamma}{q_0-q_\Gamma}\right)^n d q'} + \int_\qg^q { \delta (\hf \hrho ) dq'}\\
\nonumber
&=  \hf_0 \hrho_0 \frac{(q-\qg)^{n+1}}{(1+n)(q_0-q_\Gamma)^n} + \int_\qg^q{ \delta (\hf \hrho ) dq'}\\
&=  \hf_0  \frac{\hrho(q)(q-\qg)}{(1+n)} + \int_\qg^q{ \delta (\hf \hrho ) dq'},
\end{flalign}
where the first analytical term is precisely the contribution of the infinite gradient. The second term has finite gradient onto $\Gamma$ by construction and so, it can be estimated by numerical means without problems at second-order from the trapezoidal rule. The kernel splitting method is general and applies for computing both BVs and output values. It has no special efficiency for $n \ge 1$. It requires the knowledge of the fluid boundary.

\begin{figure}
\includegraphics[width=8.2cm,bb=25 20 330 250,clip=]{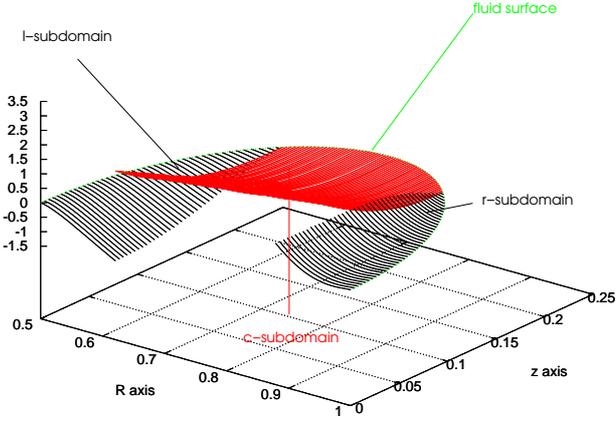}
\caption{Same legend as for Fig. \ref{fig:dsmpict.eps} but for the residue  $\delta (\hf \hrho )$ (i.e. after power-law splitting of the kernel).}
\label{fig:dsmpict_pls.eps}
\end{figure}

For the l- and r-subdomains where the radial integration is performed first, we have $q=a$, while this is $q=z$ in the c-subdomain. Note that the pivot point serves only to determine the amplitude of the power-law term through $f_0$ and must not be used in the quadrature scheme because $\delta (\hf \hrho)=0$ for both $q=\qg$ and $q=q_0$. An example of residue $\delta (\hf \hrho )$ is shown in Fig. \ref{fig:dsmpict_pls.eps}. Infinite derivatives ont $\Gamma$ have disappeared.

\subsection{Results for $n=0.5$}

\begin{figure}
\includegraphics[width=8.5cm,bb=39 46 706 522,clip=]{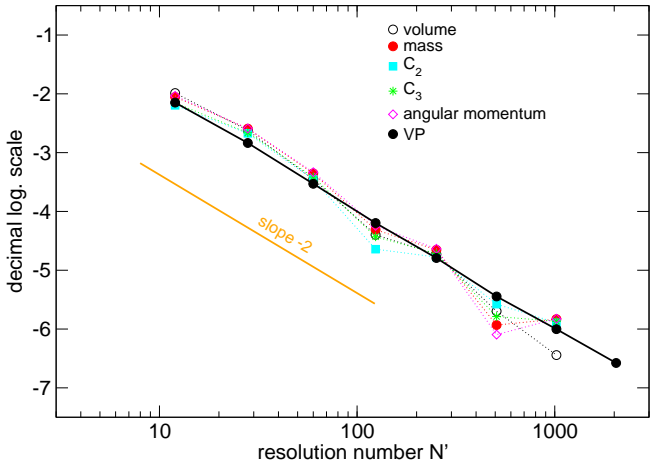}
\caption{Same legend as for Fig. \ref{fig:C2vsl_n05.eps} but with power-law splitting of the kernel.}
\label{fig:C2vsl_n05_wpls.eps}
\end{figure}

We have run the code for various resolutions and the results are given in Fig. \ref{fig:C2vsl_n05_wpls.eps}, which is to be compared to Fig. \ref{fig:C2vsl_n05.eps}. We see that the $2$nd-order accuracy is now well reproduced by using kernel splitting, for all quantities, in particular for the Virial parameter and fluid volume as well. Table \ref{tab:cpredtohachisun05} lists output quantities and compares the different approaches. While the boundary $\Gamma$ becomes explicitly an unknown of the problem, which directly impacts on BVs inside the SCF-iterations, the convergence process is not fragilized. We can still maintain a very low level of convergence close to the computer precision (see below) and still use Ineq. (\ref{eq:condh}) as the pertinent convergence criterion. The computing time is substantially increased with kernel splitting, as expected but this is not critical. 

\begin{table}
\centering
\begin{tabular}{lllll}\\
                            &                      & \multicolumn{3}{c}{|||| this work ||||}  \\
                            &                      &            & \multicolumn{2}{c}{\bf subgrid approach}\\
  quantity                  & Hachisu (1986)                &            & with $\Gamma$ & splitting\\ \hline
  $\Lambda$                 & $0.088^*$             &  $0.594$   &  $0.594$  & $0.594$\\
  $C_1$                     & ?                    & $0.1200$    & $0.1201$ & $0.1200$\\
  $C_2$ (i.e. $\Omega_0^2)$  & $0.383$              & $0.3825$   & $0.3823$ & $0.3824$\\
  $-C_3$                    &                     & $0.7729$   & $0.7725$ & $0.7729$ \\
  $i_M$          &                                & $61$        & $61$ & $61$\\
  $\hh_M$        &                                & $1$         & $1$  & $1$\\
  $\hbsur$                   &                    & $0.1581$  & $0.1583$ & $0.1582$\\
  $\hv$                      & $0.746$            & $0.7451$  & $0.7459$ & $0.7455$\\ 
  $\hm$                      & $0.471$            & $0.4710$  & $0.4707$ & $0.4710$ \\
  $\langle \hrho \rangle$    & $0.631^*$          & $0.6321$  & $0.6311$ & $0.6317$\\
  $\sqrt{C_2} \hjcin$        & $0.173$            & $0.1732$  & $0.1730$ & $0.1731$\\
  $C_2 \hcin$                & $0.536$            & $0.5354$  & $0.5349$ & $0.05353$ \\
  $-\hw$                     & $0.171$            & $0.1713$  & $0.1712$ & $0.1713$\\
  $\frac{C_1}{n+1}\hu$       & $0.0642$           & $0.6423$  & $0.6429$ & $0.06422$ \\
  $\log(VP)$                 & ?                & $-4.23$   & $-4.23$  &$-4.23$  \\
  iterations & ?        & $34$  & $33$ & $33$ \\\hline
\end{tabular}
\caption{Comparison with Hachisu (1986) for $n=0.5$ ($^*$ estimated). Results obtained by accounting for the fluid boundary $\Gamma$ are given  the implementing the kernel splitting method are listed in the third column.}
\label{tab:cpredtohachisun05}
\end{table}

\section{Acceleration of the SCF-iterations from internal preconditionning}
\label{sec:convacc}

\subsection{A recipe}

\begin{figure}
\includegraphics[width=8.4cm,bb=38 52 729 595,clip=]{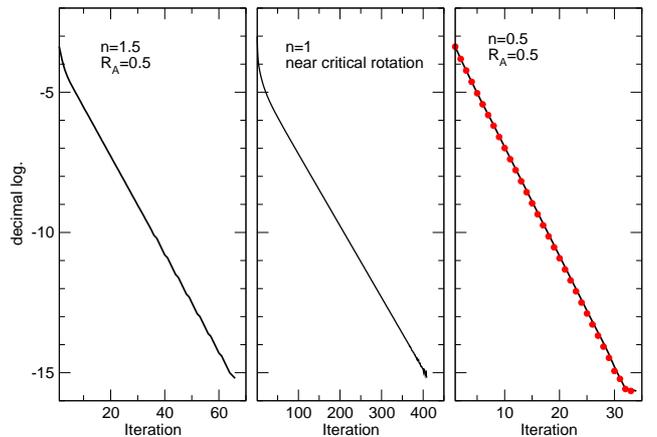}
\caption{Variation of $\dhfro$ during the SCF-iterations ({\it black}) down to convergence for the three configurations discussed in  Sect. \ref{sec:perf}. Convergence is not impacted by kernel splitting ({\it red dot}).}
\label{fig:allconv.eps}
\end{figure}

The evolution of $\dhfro$ during the SCF-iterations is shown in Fig. \ref{fig:allconv.eps} for the three configurations studied before (see Sect. \ref{subsec:scfloop}). We see the remarkable stabilisation of the SCF-iteration at a very high level of precision (see also Fig. \ref{fig:ehent_01.ps}). As mentionned, this holds whatever the polytropic index, with or without subgrid approaches. The striking feature is the exponential decay which appears after a very few iterations, once the starting guess is almost forgotten. There is obviously a sensitivity to the polytropic index and axis ratio $\hr_A/\hr_B$ of the torus, but this property appears relatively universal. The decay is not only observed for $\dhfro$, but for the enthalpy field at almost all of the grid nodes. It has an origin in the SCF-method. We can therefore try to reduce the number of iterations, at least empirically, from this decay. Let us assume that, at time $t$, we have
\begin{equation}
\ln \left| \delta \hh^{(t)} \right| \sim  a-bt.
\end{equation}
where $a>0$ and $b>0$. Since this expression also writes
\begin{flalign}
\delta \hh{(t+\tau)} &= \delta \hh^{(t+\tau-1)} e^{-b},
\end{flalign}
we can estimate field values at the end of the next iteration, i.e. at time $t+2$. With $\tau=2$, we have
\begin{flalign}
\hh^{(t+1)} + \delta \hh^{(t)}e^{-b} \equiv \hh^{(t+2), \text{estimate}}.
\label{eq:hcorrected}
\end{flalign}
Like $\hh$, $b$ is a function of space (each node has its own path to convergence), namely
\begin{equation}
b \sim - \left(\frac{d \ln \left| \delta \hh \right| }{dt}\right)_{t+1},
\end{equation}
which can be determined from finite differences. At second order, we have
\begin{flalign}
\nonumber
- b &\approx \frac{\delta \hh^{(t+\frac{3}{2})}-\delta \hh^{(t+\frac{1}{2})}}{\delta \hh^{(t+1)}}\\
         &\approx \frac{\hh^{(t+1)}-2\hh^{(t)}+\hh^{(t-1)}}{\frac{1}{2}\left[\hh^{(t+1)}+\hh^{(t-1)}\right]}.
\label{eq:bij}
\end{flalign}
The mass density field to be considered in BVs when entering step $t+1$ of the SCF-method is therefore given by
\begin{equation}
\hrho_{i,j}^{(t+1)}=\sup\left[\hh^{(t+2), \text{estimate}},0\right]^n,
\label{eq:rhocorrected}
\end{equation}
which requires the storage of the enthalpy over a $3$ successive iterations, at $t-1$, $t$ and $t+1$. 
\begin{figure}
\includegraphics[width=8.7cm,bb=76 300 561 526,clip=]{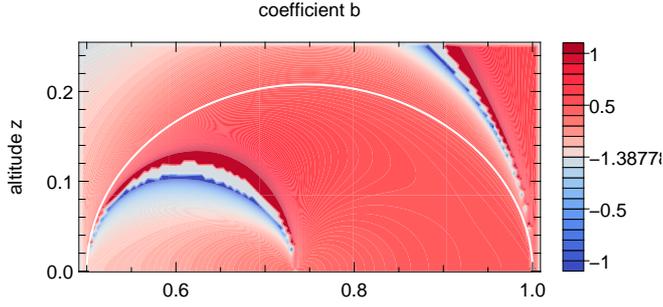}
\caption{Function $b$ computed after $5$ iterations of the classical SCF-loop for the torus considered in Sect. \ref{subsec:example1}. The criterion for preconditionning is not satisfied everywhere ({\it grey zones}). At nodes where $b>0$ ({\it red zones}), the enthalpy field has already started to converge and would be help by preconditionning.}
\label{fig:b_coefficient_at_t5.ps}
\end{figure}

The next question concerns the application of such a correction: where and when can we anticipate and provide a improved guess for the mass density at the begining of iteration $t+1$ ? We must first remind that we are mainly interested in BVs, and so only nodes of positive enthalpy are concerned. Second, the correction must not perturb the SCF-iterations too much, and cancel convergence. There is no risk at nodes where $b>0$, but this may be critical with $b < 0$. A possible criterion for preconditionning is
\begin{equation}
\begin{cases}
\hh_{i,j}>0\\
\delta \hh_{i,j}^{(t+1)} \times \delta \hh_{i,j}^{(t)} > 0,\\
b_{i,j} > b_0 \gtrsim 0,\\
\end{cases}
\label{eq:towardsc}
\end{equation}
where the second inequality means that the enthalpy field must vary monotonically at the actual node. An accelerated version of the SCF-iterations can be obtained by modifying the algorithm given in Sect. \ref{subsec:Mrecsaling} as follows

\begin{enumerate}
\setcounter{enumi}{7}
\item {\bf preconditionning (optional)}: if Ineq.(\ref{eq:towardsc}) are fulfilled, then $\rho$ is given by Eq.(\ref{eq:rhocorrected}) when entering step $t+1$, i.e. $\hh^{(t+1)}$ is replaced by $\hh^{(t+2), \text{estimate}}$.
\end{enumerate}

\subsection{Results}

\begin{figure}
\includegraphics[width=8.4cm,bb=38 52 724 583,clip=]{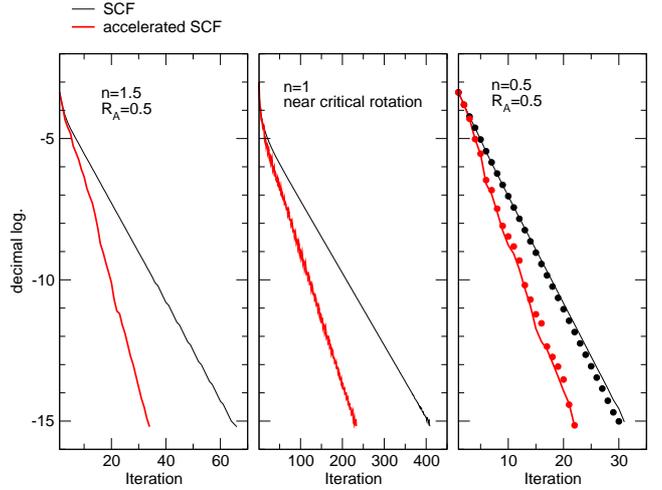}
\caption{Same legend as for Fig. \ref{fig:allconv.eps} but when preconditionning the mass density inside the cycle ({\it red}). Convergence is then attained faster, by a factor $2$ typically. This holds with a subgrid approach ({\it filled circles}).}
\label{fig:allconv_improved.eps}
\end{figure}

Figure \ref{fig:b_coefficient_at_t5.ps} displays $b$ computed from Eq.(\ref{eq:bij}) for the torus with $n=1.5$ after $5$ steps of the classical SCF-iterations. With a mean value $\langle b \rangle \sim 0.4$, the enthalpy is on the way to convergence as the deviation from one step to the other already diminishes, and preconditionning helps to accelerate the process. We have checked this recipe to many configurations and observed mostly no failure indeed. It works well with or without including boundary effects and kernel splitting for $n<1$. The results are given in Fig. \ref{fig:allconv_improved.eps} for the three examples discussed in Sect. \ref{sec:perf} where we have used $b_0=-1$. By using the same convergence criterion, the SCF-iterations require about half the normal time.

\section{Conclusion}

In this article, we have reinvestigated toroidal figures of equilibrium with a special focus on the convergence process of the SCF-iterations and accuracy of solutions. In this context, we have carried out the full problem at $2$nd-order and discussed the impact of discretisation and quadrature schemes on ouput quantities. As argued, the polytropic hypothesis sets severe constraints on the determination of solution for $n<1$ due to the non-derivability of the mass density at the fluid boundary. Though a few selected examples, we have shown how the gravitational, centrifugal and internal energy densities are distributed in space. We have demonstrated that the Virial Parameter is a good tracer of the precision of output quantities, except for geometrical quantities like the volume which fluctuates with the resolution. Such an uncertainty directly impacts on the location of critical rotation in the $\omega^2-j^2$ equilibrium digrams. Rather than increasing the order of the quadrature scheme, we have shown that the $2$nd-order can finally be recovered if the fluid boundary is detected and accounted for correctly. For polytropic indices lower than unity, quadrature troubles due to infinite gradients of the mass density can be circumvented through kernel splitting.

Finally, we have shown that it is possible to speed-up the convergence of the SCF-method by internal preconditionning of the enthalpy field. Actually, the temporal evolution of the enthalpy at the grid nodes shows an almost universal exponential decay that can be used back to improve the new guess inside the SCF-loop. This improvement is particularly interesting when building high resolution equilibrium sequences. It is a step toward the understanding of the convergence of the SCF algorithm which remains poorly understood and documented \citep{od03}. The recipe is probably not limited to toroidal structures and can probably apply to spheroidal ones, which would be interested to check.

\section*{Acknowledgments}
It is a pleasure to thank B. Boutin-Basillais. We would like to dedicate this paper to our colleague and friend Dr. J.P. Zahn.


\end{document}